\RequirePackage{ifpdf}
\documentclass[11pt,a4paper]{article}
\usepackage{epsfig,jheppubme}

\usepackage{multirow,longtable,enumerate}

\usepackage{physics}
\usepackage{tensor}
\usepackage{amsmath}
\usepackage{amssymb}
\usepackage{appendix}
\usepackage{mathtools}
\usepackage{mathrsfs}
\usepackage[normalem]{ulem}
\usepackage{tikz}
\usepackage{booktabs}
\usepackage{bbm}
\usepackage{verbatim}
\usepackage{comment}
\usepackage{xspace}
\usepackage{longtable}

\usepackage{graphicx}
\usepackage{amsthm}
\usepackage{slashed}
\usepackage{hyperref}
\usepackage{setspace}
\usepackage{tikz}
\usepackage{array}
\usetikzlibrary{shapes}

\usepackage[tableposition=below]{caption}
\font\cmss=cmss12

\newcommand\bi{\begin{itemize}}
\newcommand\ei{\end{itemize}}

\newcommand\bmbA{{\overbar{\mathbf A}}}
\newcommand\mbA{{\mathbf A}}

\newcommand\bea{\begin{eqnarray}}
\newcommand\eea{\end{eqnarray}}
\newcommand\be{\begin{equation}}
\newcommand\ee{\end{equation}}

\newcommand\btau{{\bar \tau}}
\newcommand\bq{{\bar q}}

\newcommand\cN{{\cal N}}

\newcommand\bchi{{\overline \chi}}

\newcommand\sfrac[2]{{\textstyle\frac{#1}{#2}}}

\newcommand\ZZ{\hbox{Z\kern-.4emZ}}
\newcommand\sZZ{\hbox{\sevenfont Z\kern-.4emZ}}

\newcommand{\eref}[1]{Eq.\,(\ref{#1})}
\newcommand{\adst}{AdS$_3$\xspace}
\newcommand{\poin}{Poincar\'e\xspace}
\newcommand{\poinsum}{Poincar\'e sum\xspace}
\newcommand{\poinser}{Poincar\'e series\xspace}
\newcommand{\sutk}{SU(2)$_k$\xspace}
\newcommand{\suthk}{SU(3)$_k$\xspace}
\newcommand{\sunk}{SU($N$)$_k$\xspace}
\newcommand{\sunone}{SU($N$)$_1$\xspace}

\newcommand{\overbar}[1]{\mkern 1.5mu\overline{\mkern-1.5mu#1\mkern-1.5mu}\mkern 1.5mu}

\newcommand{\bc}{\mathbf{c}}

\newcommand{\gtau}{\tau_\gamma}
\newcommand{\bgtau}{{\bar\tau}_\gamma}

\allowdisplaybreaks

\def\IB{\relax{\rm I\kern-.18em B}}
\def\IC{{\relax\hbox{\kern.3em{\cmss I}$\kern-.4em{\rm C}$}}}
\def\ID{\relax{\rm I\kern-.18em D}}
\def\IE{\relax{\rm I\kern-.18em E}}
\def\IF{\relax{\rm I\kern-.18em F}}
\def\II{\relax{\rm I\kern-.18em I}}
\def\IZ{\mathbb{Z}}
\def\Id{\relax{1\kern-.32em 1}}
\def\IG{\relax\hbox{$\inbar\kern-.3em{\rm G}$}}
\def\IR{\relax{\rm I\kern-.18em R}}

\allowdisplaybreaks

\title{Poincar\'e Series, 3d Gravity and Averages of Rational CFT} 

\author[a]{Viraj Meruliya\,}
\author[a]{Sunil Mukhi\,}
\author[b]{and Palash Singh\,}

\affiliation[a]{Indian Institute of Science Education and Research,\\  Homi Bhabha Rd, Pashan, Pune 411 008, India}

\affiliation[b]{Mathematical Institute, University of Oxford,\\ Woodstock Road, Oxford, OX2 6GG, United Kingdom} 

\emailAdd{viraj.meruliya29@gmail.com}
\emailAdd{sunil.mukhi@gmail.com}
\emailAdd{palash13singh@gmail.com}

\abstract{We investigate the \poinser approach to computing 3d gravity partition functions dual to Rational CFT. For a single genus-1 boundary, we show that for certain infinite sets of levels, the SU(2)$_k$ WZW models provide unitary examples for which the \poinser is a positive linear combination of two modular-invariant partition functions. This supports the interpretation that the bulk gravity theory (a topological Chern-Simons theory in this case) is dual to an average of distinct CFT's sharing the same Kac-Moody algebra. We compute the weights of this average for all seed primaries and all relevant values of $k$. We then study other WZW models, notably SU($N$)$_1$ and SU(3)$_k$, and find that each class presents rather different features. Finally we consider multiple genus-1 boundaries, where we find a class of seed functions for the \poinsum that reproduces both  disconnected and connected contributions -- the latter corresponding to analogues of 3-manifold ``wormholes'' --  such that the expected average is correctly reproduced. 
}
\preprint{}


\arxivnumber{}

\begin{document}

\maketitle

\section{Introduction and background}

The problem of computing the partition function of pure Einstein gravity with \adst boundary conditions was addressed in \cite{Maloney:2007ud} using the method of \poin sums (earlier work on these sums in related contexts can be found in \cite{Dijkgraaf:2000fq}, in this context see also \cite{Manschot:2007zb, Manschot:2010mod}). One considers Euclidean gravity on manifolds whose asymptotic boundary is a torus and assumes a semi-classical summation formula. This is physically motivated by the existence of an SL(2,$\mathbb{Z}$) family of Euclidean black holes, all of which have a torus boundary, that are assumed to be the relevant saddle-points of a gravity path integral. The result is a \poinsum whose first term is the contribution from the ``vacuum'' spacetime, namely Euclidean \adst, while the remaining terms are modular transforms of this contribution. 

Subsequently the results of \cite{Maloney:2007ud} were streamlined and generalised in  \cite{Giombi:2008vd, Keller:2014xba} among other works. One of the important generalisations was the fact that the \poin series need not start with the vacuum contribution. Instead one can start with the partition function for any ``seed'' primary and compute the \poin sum over this. More generally the contributions from different seeds can be added, leading to a large family of candidate \adst gravity partition functions. Unfortunately the results in \cite{Maloney:2007ud, Keller:2014xba} have certain undesirable features that have proved very difficult, if not impossible, to eliminate even allowing for such generalised seeds.

Following  \cite{Maloney:2007ud}, it was proposed in \cite{Castro:2011zq} to examine whether a similar \poin series approach could be used with a different starting point: the existence of unitary minimal-model RCFT's with central charge $\bc<1$ and the fact that their modular-transformation matrices $\gamma(a,b,c,d)$ are known explicitly. One may define the gravity duals of such theories, though they will not be semi-classical, by treating the identity character of the RCFT as a ``seed'' primary and assuming its partition function to be the analogue of a ``vacuum contribution'' to a gravity partition function. The rest of the partition function is then obtained by summing a \poinser, which in this case is a finite sum. The result of \cite{Castro:2011zq} is essentially a no-go statement about unitary $(m,m+1)$ minimal models: beyond the first two models, the gravity partition functions are not equal to CFT partition functions. Instead one finds linear combinations of physical CFT partition functions in a few cases, and for many values of $m$ one also gets linear combinations that include unphysical modular invariants (in the sense of having negative coefficients in their $q,\bq$ expansion). Thus, it was concluded that most unitary minimal models cannot be interpreted as being dual to \adst gravity\footnote{More recently it has been argued in \cite{Jian:2019ubz} that studying the \poinsum approach with genus-two boundaries provides additional constraints that exclude even the second ($m=4$) minimal model.}.

In recent years there has been a realisation that in low dimensions, AdS gravity may be dual not to a single CFT but to a weighted ensemble of CFT's \cite{Saad:2019lba, Witten:2020wvy, Maloney:2020nni, Afkhami-Jeddi:2020ezh}. As briefly noted in \cite{Maloney:2020nni}, the result of \cite{Castro:2011zq} for unitary minimal models might be an example of the same phenomenon. This motivates us to examine the \poinsum for various classes of RCFT and try to understand whether it can be interpreted in terms of either a single RCFT dual or an average over finitely many RCFT's. In the former case one can claim a (non-semi-classical) AdS/CFT duality, while in the latter one can try to interpret the dual as a weighted average over different physical CFT's. The latter interpretation requires in particular that the weights appearing in the average be non-negative, so one has to look for classes where this is the case. Next, one may go beyond the case of a single toroidal boundary since additional consistency conditions can arise by considering 3-manifolds with multiple disconnected torus boundaries, as well as higher-genus boundaries. In particular, with multiple genus-1 boundaries the interpretation as an average of CFT's should work with the same weights as for single boundaries. We will find examples that satisfy all  of these requirements\footnote{We will not consider higher-genus boundaries in this paper and will only comment briefly on this at the end.}. We expect that this study could teach us valuable lessons, in relatively simple cases, of how averaging over CFT's works in the AdS/CFT correspondence. 

Compared with the general (irrational) case, the \poinsum for RCFT is a finite sum and has to be performed over a coset $\Gamma_c\backslash \Gamma$ where $\Gamma=$ SL(2,$\mathbb{Z}$) and $\Gamma_c$ is the finite-index subgroup that keeps the seed contribution to the partition function invariant. This needs to be worked out case by case and quickly becomes quite tedious. Also the precise definition of $\Gamma_c$ depends on the seed rather than being just a property of the characters. Hence we resort to a more practical approach: we carry out our \poinsum over the coset $\Gamma(N_0)\backslash \Gamma$ where $\Gamma$ is the principal congruence subgroup of SL(2,$\mathbb{Z}$), while $N_0$ is an integer that we will define (it is similar, but not equal, to the ``conductor'' $N$ of an RCFT defined in \cite{Bantay:2001ni}). This is a single group for each set of characters, independent of which primary is used as the seed, and is easily found just be evaluating $N_0$. It is a subgroup of the true invariance group $\Gamma_c$ of the given seed partition function, and since it has finite index in $\Gamma_c$ (which in turn has finite index in $\Gamma$) we will just get the desired result of the \poinsum with an overall factor that can be removed by a normalisation. 

For the \sunk WZW models the criteria for which one finds different numbers of RCFT in the sum depend on $N$ and the level $k$. Depending on these values we find precise criteria under which the \poinsum yields the partition function of a single RCFT or a linear combination of physical RCFT's. There are infinitely many values of $k$ for which each of these possibilities is realised. Where the \poinsum yields a sum of distinct partition functions, we will find a method to compute the weights of this average as a function of $N,k$, and work it out in the simplest cases where the average is over two CFT's. We will work things out in considerable detail for the unitary \sutk WZW models, and subsequently discuss aspects of \sunk for $N\ge 3$ \footnote{While discussing the ``gravity duals'' of WZW models, we will make some tentative comments on what this should mean.}. We also revisit the minimal (unitary and non-unitary) series, for which the analysis is quite similar to \sutk in many aspects.

We then move on to the \poinsum for multiple genus-1 boundaries. We will find that for carefully chosen seeds, one finds answers such that the partition function is an average over products of CFT partition functions with the same weights, for any number of boundaries. In the irrational case this type of setup implies the existence of ``wormhole'' contributions from manifolds with multiple boundaries \cite{Saad:2018bqo, Saad:2019lba, Cotler:2020ugk, Cotler:2020hgz} and we propose analogous ``wormhole'' seeds in the RCFT context. 

Let us now briefly review the approach of \cite{Maloney:2007ud, Keller:2014xba} (henceforth referred to as MWK) to the calculation of the semi-classical partition function of AdS$_3$ Einstein gravity, as well as its generalisation to the RCFT case in \cite{Castro:2011zq}. The MWK approach sums a seed over a set of manifolds denoted $M_\gamma$ where $\gamma\in \Gamma=$ SL(2,$\mathbb{Z}$). Of these, $M_{\mathbf 1}$ is thermal \adst and the others are Euclidean BTZ black holes. The most obvious starting point is to take thermal \adst as the seed. Because the BTZ black holes are SL(2,$\mathbb{Z}$) transforms of thermal \adst, we have:
\be
Z_{\gamma}(\tau,\btau)=Z_{\mathbf 1}(\gtau,\bgtau)
\ee
where:
\be
\gamma=\begin{pmatrix}
a & b \\ c & d
\end{pmatrix}\in \Gamma,\qquad 
\gtau=\frac{a\tau+b}{c\tau+d}
\ee
Thus one would naively write the gravity partition function as a \poinsum:
\be
Z(\tau,\btau)=\sum_{\gamma\in\Gamma}Z_{\gamma}(\tau,\btau)=\sum_{\gamma\in\Gamma}Z_{\mathbf 1}(\gtau,\bgtau)
\label{idseed}
\ee
However, SL(2,$\mathbb{Z}$) elements of the form:
\be
\gamma=\begin{pmatrix}
1 & n \\ 0 & 1
\end{pmatrix}
\ee
leave $Z_{\mathbf 1}$ invariant, and these generate a group isomorphic to $\IZ$. So the sum above is replaced by a \poinsum over coset representatives:
\be
Z(\tau,\btau)=\sum_{\gamma\in\IZ\backslash \Gamma}Z_{\mathbf 1}(\gtau,\bgtau)
\ee
The rest of the calculation proceeds by evaluating $Z_{\mathbf 1}$ using the action and symmetries of thermal \adst, and then evaluating the sum. After regularisation the result is finite but unfortunately has several undesirable features including negativity of coefficients that should correspond to degeneracies of states. A relatively recent work \cite{Benjamin:2019stq} noted a large-scale occurence of negative coefficients in the $q$-series. One class of remedies that has been proposed \cite{Keller:2014xba, Benjamin:2020mfz, Alday:2019vdr} is to cure the negativity by generalising from the identity seed to a sum over suitably selected primary seeds. A different remedy \cite{Maxfield:2020ale} is based on a relation (via Kaluza-Klein reduction) to the analogous problem for AdS$_2$, which suggests one should sum over additional 3-manifolds -- however the compatibility of this proposal with modular invariance is not yet clear. 

To find gravity duals for rational CFT, the approach pioneered in \cite{Castro:2011zq}, starts from a slightly different perspective. This work considers the unitary Virasoro minimal models with central charge $0<\bc<1$. At such small values of $\bc$ the gravity dual (if such a concept has a meaning) would have an AdS radius $\ell_{\rm AdS}=\frac{2G_N \bc}{3}\sim G_N$ which is far from the semi-classical regime. So one gives up the idea of summing over manifolds that are classical solutions. In its place, one considers the partition function $Z_{\rm Id}(\tau,\btau)=|\chi_{\rm Id}|^2$ coming from secondaries over the identity primary (after null vectors are removed). Defining $\Gamma_c\subset\Gamma$ as the subgroup of $\Gamma$ that preserves $Z_{\rm Id}$, one evaluates the \poinsum:
\be
Z(\tau,\btau)=\sum_{\gamma\in\Gamma_c\backslash \Gamma}Z_{\rm Id}(\gtau,\bgtau)
\label{psum}
\ee
Unlike the semi-classical case this sum is finite, because for RCFT $\Gamma_c$ is a finite-index subgroup of $\Gamma$ and therefore the coset has a finite number of representative elements. 

When applied to the $(m,m+1)$ Virasoro minimal models, the results of \cite{Castro:2011zq} are as follows. For $m=3,4$ there are unique modular invariants built out of the characters, and the \poinser \eref{psum} reproduces these up to an overall factor (this is the unique possibility consistent with modular invariance). But for $m\ge 5$ the \poinser for the identity seed leads, in the first few cases, to a linear combination of the allowed physical CFT partition functions \cite{Cappelli:1986hf, Cappelli:1987xt}. At larger values of $m$ one starts to encounter unphysical partition functions (those whose coefficients in the $q,\bq$ series are either non-integral when the identity state is normalised to unity, or non-positive, or both) and these appear in the \poinsum \footnote{We note in passing that the existence of modular-invariant partition functions with negative integral coefficients seems to be a non-holomorphic version of the ``quasi-characters'' extensively studied in \cite{Chandra:2018pjq, Chandra:2018ezv, Mukhi:2020gnj}. It seems that negative integrality is a rather widespread feature in the context of both holomorphic vector-valued modular forms and non-holomorphic modular invariants.}.
 
In what follows, we examine the \poin series approach for rational CFT, focusing primarily on \sunk WZW models but also revisiting the case of Virasoro minimal models. In Section \ref{genus1} we explicitly evaluate the \poinsum for a number of models, working successively with \sutk, \sunone and \suthk. For the first two cases we develop an approach to compute the weights of the resulting average over CFT's as a function of $k, N$ respectively and identify infinite families of models (and seed primaries) for which these weights are non-negative. Thereafter we revisit the minimal models and compute the weights of the average for cases where there are two physical invariants. In Section \ref{highergenus} we study the \poin series for multiple genus-1 boundaries, focusing on \sutk. The conclusions are in Section \ref{conclusions}, while the Appendices provide some basic data on the relevant CFT's as well as some additional tables with the results of our calculations. 

\section{\poin Sums for a Single Genus-1 Boundary}

\label{genus1}

\subsection{Generic RCFT}

\label{psumgeneric}

Let us start by defining the \poinsum for an arbitrary RCFT. We start with a set of holomorphic characters $\chi_0(\tau),\chi_1(\tau),\allowbreak \cdots,\chi_{P-1}(\tau)$. These form a vector-valued modular form (VVMF) under $\Gamma =$ SL(2,$\mathbb{Z}$). Thus, under a modular transformation:
\be
\gamma=
\begin{pmatrix}
a & b\\ c& d
\end{pmatrix}
\in \Gamma
\ee
we have:
\be
\tau\to \tau_\gamma=\frac{a\tau+b}{c\tau+d}
\ee
and:
\be
\chi_i(\tau_\gamma)=\sum_{j=0}^{P-1} (M_{\gamma})_{ij}\chi_j(\tau) 
\ee
for some matrix $M_\gamma$. The above relation can be written more briefly as $\chi(\tau_\gamma)=M_\gamma\,\chi(\tau)$.

Each character has the behaviour:
\be
\chi_i(\tau)=q^{\alpha_i}\big(a_0^{(i)}+a_1^{(i)}q+a_2^{(i)}q^2+\cdots\big)
\label{charexp}
\ee
for some rational exponents $\alpha_i$ and non-negative integers $a_n^{(i)}$. In a unitary theory, $\alpha_0$ is the smallest of the $\alpha_i$ and is identified with $-\frac{\bc}{24}$ where $\bc$ is the central charge. The remaining $\alpha_i$ are then identified with $h_i-\frac{\bc}{24}$ where $h_i$ are the holomorphic conformal dimensions.  

Now we pick one of the characters, not necessarily the identity, and consider:
\be
Z_{ii}(\tau,\btau)\equiv|\chi_i(\tau)|^2
\label{diagseed}
\ee
We think of this as the partition function corresponding to the ``spinless seed primary'' $i$ of dimension $(h_i,h_i)$. It is not modular-invariant by itself. We can also allow non-diagonal primaries with dimensions $(h_i,h_j)$ for $j\ne i$ as long as $h_j-h_i$ is an integer. The corresponding seed partition function would be:
\be
Z_{ij}(\tau,\btau)\equiv\bchi_i(\btau)\chi_j(\tau)+{\rm cc}
\label{nondiagseed}
\ee
Next we take $Z_{\rm seed}$ to be any of the various seed primaries as above, or a linear combination of them. The relevant \poinser is then:
\be
Z(\tau,\btau)\equiv \sum_{\gamma\in \Gamma_c\backslash \Gamma}Z_{\rm seed}(\gtau,\bgtau)
\label{genpsum}
\ee
where $\Gamma_c$ is the subgroup of $\Gamma$ that leaves $Z_{\rm seed}$ invariant.

The result of the sum in \eref{genpsum} is manifestly modular invariant.  Thus it must be a linear combination of all allowed modular invariants for the given set of characters. A basis for these is:
\be
 Z_J(\tau,\btau)= \bchi(\btau)I_J\,\chi(\tau)
\equiv\sum_{i,j=0}^{P-1}\bchi_i (I_J)_{ij}\chi_j
\label{avgc}
\ee
were $I_J$ is a basis of matrices that lead to modular-invariants $Z_J$ and $J$ runs over the number of independent matrices of this type. It follows that:
\be
Z(\tau,\btau)=\sum_J c_J Z_J(\tau,\btau)
\ee
where $c_J$ are some basis-dependent coefficients. Such invariants sometimes have all positive coefficients in their $q,\bq$ expansion. Then they can correspond to the partition function of a CFT, but only if they satisfy some additional requirements such as having a unique vacuum state. In this case we refer to them as ``physical''. Alternatively they may have some negative Fourier coefficients, in which case they cannot correspond to a CFT. 

Thus the possible results of performing a \poinsum are of three types: (I) when the index $J$ only takes one value and the invariant corresponds to a physical RCFT, we get the partition function of that RCFT, (II) when the index $J$ ranges over multiple modular invariants that all correspond to physical RCFT's, we get a linear combination of partition functions of these different RCFT's, (III) when the index $J$ runs over both physical and unphysical invariants, we get a result that cannot be interpreted as an average over RCFT. Within type (II) there are two sub-classes: if all the $c_J\ge 0$, the linear combination can be interpreted as a weighted average. This case can be called type (IIa). On the other hand if at least one of the $c_J$ is negative then the averaging interpretation is no longer tenable, this is called (IIb).  This classification applies independently for any $Z_{\rm seed}$, which, as indicated above can be associated to a single seed primary or a sum over seed primaries.
 
In view of our discussion above, we will consider types (I) and (IIa) to be physically acceptable, with the latter corresponding to an average over RCFT's with definite weights, while types (IIb) and (III) will be rejected as unphysical . Our primary goal is to explore the type (IIa) behaviour in considerable detail in many examples, as well as to find general formulae for the coefficients $c_J$ in families of theories where such behaviour holds. 

It remains to find the appropriate coset to sum over when performing a \poinsum for RCFT. As indicated above, it is sufficient to search for a finite-index subgroup of $\Gamma$ that leaves the seed partition function $Z_{\rm seed}$ invariant. This may be smaller than the largest possible such group, that we have earlier called $\Gamma_c$, but this does not matter since the result for the \poinsum will be the same up to an overall factor.

We need a few definitions. The {\em conductor} $N$ of an RCFT is defined as the LCM of the denominators of the exponents $\alpha_i$ defined in \eref{charexp} above. Then, it is well-known \cite{Bantay:2001ni} that the principal congruence subgroup $\Gamma(N)$ leaves each character separately invariant\footnote{Various congruence subgroups, among which we will make use of $\Gamma(N)$ and $\Gamma_1(N)$ in particular, are defined in Appendix \ref{congsub}.}:
\be
\chi_i(\gtau)=\chi_i(\tau),\quad \gamma\in \Gamma(N)
\ee
In general, $\Gamma(N)$ is a proper subgroup of the kernel of the modular representation (the largest subgroup of $\Gamma$ that preserves the characters). 

However, this result is too strong for us because we are interested in the subgroup that preserves $Z_{ij}=\bchi_i\chi_j$ (where $|h_i-h_j|$ is integer) rather than $\chi_i$ itself. For this, we my consider the larger subgroup $\Gamma(N_0)$ where $N_0$ is defined as follows. Picking $\alpha_0$ to be the most negative exponent (it will eventually be identified with $-\frac{\bc}{24}$) we define the conformal dimensions associated to our characters by $h_i=\alpha_i-\alpha_0$. Now we define $N_0$ to be the LCM of the denominators of the $h_i$. We will refer to $N_0$ as the {\em semi-conductor}. 

All elements of $\Gamma(N_0)$ preserve each character up to an overall phase, which implies that $\Gamma(N_0)$ is contained in $\Gamma_c$. The argument is as follows. The proof of \cite{Bantay:2001ni} that $\Gamma(N)$ preserves all characters makes use of the standard action of the generators $T,S$ of SL(2,$\mathbb{Z}$) on the characters. Of these,
$T$ acts as a phase:
\be
T:\quad \chi_i\to e^{2\pi i\alpha_i}\chi_i  
\ee
where $\alpha_i=-\frac{\bc}{24}+h_i$. Now we can modify the $T$ generator by defining:
\be
T'=\omega T 
\ee
where $\omega=e^{\frac{2\pi i \bc}{24}}$. Then $T'$ and $S$ also form a representation of SL(2,$\mathbb{Z}$), and we can find the subgroup that preserves the characters under this action. By the same arguments as before, this group will be $\Gamma(N_0)$. It follows that the subgroup that preserves characters upto a common phase, in the usual action, is $\Gamma(N_0)$. Since the seed partition function is blind to this phase, $\Gamma(N_0)$ is therefore a subset of $\Gamma_c$. It may be noted that $N_0$ divides the conductor $N$, and therefore $\Gamma(N_0)$ is a superset of $\Gamma(N)$. 

For diagonal seeds of the form $Z_{ii}=|\chi_i|^2$ one can find a larger invariance group than $\Gamma(N_0)$. For this one uses the fact that $T$ is not an element of $\Gamma(N_0)$. However it multiplies each character by a (distinct) phase, so it preserves every seed partition function. $T$ generates a hyperbolic subgroup of $\Gamma$, of which the independent elements that are not in $\Gamma(N_0)$ are $T, T^2,\cdots, T^{N_0-1}$. We can now combine $T$ with $\Gamma(N_0)$ and find the resulting group, which turns out to be $\Gamma_1(N_0)$ (this is defined in Appendix \ref{congsub}). This follows from the inclusions $\Gamma(N_0)\subset \Gamma_1(N_0)$ and $T^n\in \Gamma_1(N_0)$, from which we see that the combination has at least $N_0$ times the number of elements of $\Gamma(N_0)$. Because $ [\Gamma_1(N_0):\Gamma(N_0)]=N_0$ it also has at most this number of elements, therefore it must be precisely the congruence subgroup $\Gamma_1(N_0)$. 

In what follows, we will work with both $\Gamma(N_0)$ and $\Gamma_1(N_0)$ at different points. The former group has the advantage of being normal, so left- and right-cosets are the same. Some of the calculations are also simpler. However the latter is a larger group and therefore has a smaller index in $\Gamma$. As a result the calculation of the \poinsum has fewer terms and this economy is helpful when working with larger values of $N_0$. 

 Thus, we finally define the Poincar\'e sum for an RCFT as:
\be
Z(\tau,\btau)\equiv \frac{1}{\cN}\sum_{\gamma\in \Gamma_{\rm sub}\backslash \Gamma}Z_{\rm seed}(\gtau,\bgtau)
\label{GPsum}
\ee
where $\Gamma_{\rm sub}$ will be either $\Gamma(N_0)$ or $\Gamma_1(N_0)$. In each case this will be a finite sum. Here $\cN$ is a normalisation factor such that the result has a leading term normalised to unity. Though our notation does not show this explicitly, the normalisation factor will depend on the seed. Note that depending on the choice of $\Gamma_{\rm sub}$, the values of the coefficients $c_I$ in \eref{avgc} will be scaled by an overall constant. 

Let us now describe the method to derive the coefficients $c_J$ (this was used in \cite{Castro:2011zq} for specific minimal models). Consider a general seed for the \poinser. This can be written as:
\begin{equation}
Z_{\rm seed}(X,\tau,\btau) =\bchi(\btau) \,X_{\rm seed}\,\chi(\tau)
\end{equation}
where $X_{\rm seed}$ is a $P \times P$ matrix. Now under a modular transformation we have:
\begin{equation}
Z_{\rm seed}(X,\gtau,\bgtau) = \bar{\chi}(\bgtau)\, X_{\rm seed} \, \chi(\gtau) = \bar{\chi}(\btau)\, (M^{\dagger}_{\gamma}X_{\rm seed}M_{\gamma}) \,\chi(\tau)
\end{equation}
The result of performing a Poincar\'e sum over $Z_{\rm seed}$ will be denoted $Z(X,\tau,\btau)$ to highlight its dependence on the seed $X$ (this is the modular invariant partition function that in other contexts we just called $Z(\tau,\btau)$). Then:
\begin{equation}
Z(X,\tau,\btau)= \sum_{\gamma}  Z_{\rm seed}(X,\gtau,\bgtau) = \bar{\chi}(\btau)\, \Big(\sum_{\gamma} M^{\dagger}_{\gamma}X_{\rm seed}M_{\gamma}\Big) \, \chi(\tau)
\end{equation}
Thus, the modular-invariant matrix $\sum_{\gamma}M^{\dagger}_{\gamma}X_{\rm seed}M_{\gamma}$ determines what linear combinations of invariants will appear in the final answer. Referring to \eref{avgc}, we have:
\begin{equation}
\sum_{\gamma}M^{\dagger}_{\gamma}X_{\rm seed}M_{\gamma} = \sum_{J=1}^D \,c_{J}\,I_{J}
\end{equation}

Now, we can define an inner product on the space of matrices as:
\be
d_{JK}=\Tr(I_JI_K)
\label{dmat}
\ee
Using this in the above equation, we have:
\begin{equation}
\begin{split}
\sum_{J}  d_{KJ}\,c_J &= \sum_{\gamma} \mbox{Tr}(I_{K} M^{\dagger}_{\gamma}X_{\rm seed}M_{\gamma}) \\
&= \sum_{\gamma} \mbox{Tr}(I_{K} X_{\rm seed}) \\
&= \left|\Gamma_{\rm sub}\backslash \Gamma\right| \, \mbox{Tr}(I_{K} X_{\rm seed}) 
\end{split}
\label{dmateqn}
\end{equation}
where in the intermediate step we used modular invariance of the $I_J$. This
matrix equation can be inverted to give:
\begin{equation}
c_{J} = \left|\Gamma_{\rm sub}\backslash \Gamma\right|\, \sum_K d^{-1}_{JK}\Tr (I_{K}X_{\rm seed})
\label{msolve}
\end{equation} 
These are the desired coefficients which, after normalisation, should be interpreted as the weights with which we average different CFT's.

\subsection{SU(2)$_k$ WZW Models}

We now show that in the unitary \sutk WZW models there are infinitely many realisations of both type (I) and type (IIa) behaviour as described in the previous section. In the latter case we will be able to compute the coefficients $c_J$ for generic models within such families and classify the cases where they are non-negative.

Before embarking on the calculation let us briefly remark on the generalisation from Virasoro to current-algebra minimal models. By generic rules of the AdS/CFT correspondence, this  should mean that the corresponding gravity theory will have gauge fields for the corresponding Lie algebra, in this case SU(2). So at first one might think the gravity theory should be Einstein gravity coupled to a Yang-Mills field. However, it has been suggested in \cite{Maloney:2020nni} in a related case, that where a current algebra is present in the CFT, the dual theory could be {\em pure} Chern-Simons theory. The logic is that in the boundary theory the energy-momentum tensor  is not an independent field but a composite of the currents via the Sugawara construction, hence the bulk should have no independent metric field, but only gauge fields that are the counterparts of boundary currents. Since the bulk theory is topological we do have a general coordinate invariant theory, something like a gravity theory without gravitons. This in turn is not so strange, given that \adst Einstein gravity anyway has no local graviton degrees of freedom in the bulk. 

Applying that logic here, the bulk dual for the \sunk WZW model should be two copies of pure SU($N$) Chern-Simons theory, one representing left-movers and the other representing right-movers of the CFT. Denoting the gauge fields by matrix valued 1-forms as $\mbA=\mbA_\mu\, dx^\mu, \bmbA=\bmbA_\mu\, dx^\mu$, the action is then:
\be
S_{\rm bulk}=\frac{k}{4\pi} \left(\int \tr(\mbA\wedge d\mbA+\sfrac23 \mbA\wedge\mbA\wedge\mbA)-
\int \tr(\bmbA\wedge d\bmbA+\sfrac23 \bmbA\wedge\bmbA\wedge\bmbA)
\right) 
\ee
This difference action is well-known to lead to a parity-conserving theory with an SU($N$)$_k \times$ SU($N$)$_k$ current algebra on its boundary, and has found applicability in the context of multiple membranes in M-theory as well as more generally in the context of a novel Higgs mechanism in 3d (for a review and references to the original works, see \cite{Bagger:2012jb}). The novelty here is that we are not simply considering the path integral on a fixed 3-manifold, but performing a \poinsum that is to be thought of as the analogue (for small $\ell_{\rm AdS}$) of a sum over 3-manifolds.

Now we turn to a discussion of $SU(2)_{k}$ WZW models in the present context. The basic features of the models are summarised in Appendix \ref{basicdata} and we can perform the Poincar\'e sums using this data. The main information required is the form taken by the $S$ and $T$ matrices when acting on the characters $\chi_\lambda$ (note that we are now labelling the characters by $\lambda=2j+1$, the multiplicity of the isospin $j$ representation, which goes from 1 to $k+1$). 

We will find that the prime factors of the shifted level $n=k+2$, also known as the ``height'', determine the behaviour of the \poinsum in the corresponding model. 
It is easily verified that the semi-conductor, defined in Section \ref{psumgeneric}, takes the simple form $N_{0} = 4n$, except for $k=1$ in which case $N_{0}=4$. We started by calculating the Poincar\'e sums for the seed corresponding to each individual character for the values $k=1,2,\dots,41$ using SageMath \cite{Sage}. The results are given in Table \ref{SU2fulltable}  in Appendix \ref{sagecalc}.

In general the results of the \poinsum are linear combinations of multiple modular invariants. As shown in \cite{Cappelli:1986hf, Cappelli:1987xt}, these consist of the following physical invariants: $Z_A$ which exists for all $n$, $Z_D$ which exists for all even $n$, and $Z_E$ which exists for $n=12, 18, 30$, in addition to unphysical invariants that exist for various values of $n$. Where our answer involves only physical invariants we write it as $c_AZ_A+c_DZ_D+c_EZ_E$ where any one or two of the coefficients $c_A,c_D,c_E$ may vanish. These coefficients are only defined up to an overall normalisation in view of the fact that we summed over cosets of $\Gamma$ by a group $\Gamma(N_0)\subset\Gamma_c$.  

We make the following observations about the results in this Table:
\begin{itemize}
\item
Within a given model, several distinct seeds give the same linear combination of invariants.

\item The vacuum primary seed always leads to a linear combination involving non-negative integers unless there are unphysical invariants at the given level. 

\item For $n=4\rho, (\rho \ge 2)$, the ratio of coefficients $\frac{c_{D}}{c_{A}} = 1$ for the vacuum seed.

\item For $n=4\rho+2, (\rho \ge 1)$, the ratio of coefficients $\frac{c_{D}}{c_{A}} = -\frac{1}{2}$ for the seed $\chi_{2}$.

\item
For certain values of $n$ and for specific seeds, one finds only $Z_A$ or only $Z_D$ as the answer. 

\end{itemize}

We will now explain these observations and generalise them to arbitrary values of $n=k+2$. For this  we briefly review some details of the results of \cite{Cappelli:1986hf, Cappelli:1987xt}. Modular invariants made of the SU(2)$_k$ characters at any fixed $k$ can be written as follows:
\begin{equation}
Z_{\delta} = \frac{1}{2}\sum_{\lambda,\lambda'=1}^{2n} \bar{\chi}_{\lambda}\, \left(\Omega_{\delta}\right)_{\lambda\lambda'} \, \chi_{\lambda'}
\label{Zdsutwo}
\end{equation}
where $\delta$ is a positive integer that divides $n$. The range of indices on the characters and the matrices $\Omega_{\delta}$ has been doubled, with a constraint: they are defined modulo $2n$ and obey the relations $\chi_{\lambda} = \chi_{\lambda + 2n} = -\chi_{-\lambda}$. These relations imply in particular that $\chi_{n}=0$. 

The matrices $\Omega_{\delta}$ are as follows. For a given $\delta$ we define $\alpha = [\delta,n/\delta]$ (the GCD of the two integers) and $\omega^{2} = 1$ mod $\frac{4n}{\alpha^{2}}$. Then:
\begin{equation}
\left( \Omega_{\delta} \right)_{\lambda\lambda'} = \begin{cases} 0, \hspace{3.4cm} \alpha \nmid \lambda \mbox{ or } \alpha \nmid \lambda'    \\ \sum_{\xi=0}^{\alpha-1} \delta_{\lambda', \, \omega\lambda + 2\xi n/\alpha}, \quad \mbox{otherwise} \end{cases}
\label{Omegasutwok}
\end{equation}
Not all the $Z_{\delta}$ are linearly independent. For example, for  $\delta=n$ we have  $\alpha=1,\, \omega=1$, so $(\Omega_{n})_{\lambda\lambda'} = \delta_{\lambda\lambda'}$, while for $\delta=1$ we have $\alpha=1,\, \omega=-1$, so $(\Omega_{1})_{\lambda\lambda'} = \delta_{\lambda,-\lambda'}$. This means that $Z_{n} = -Z_{1}$. In general, $\Omega_{\delta}$ and $\Omega_{n/\delta}$ will have the same value of $\alpha$ but $\omega$ changes sign to $-\omega$. Hence the non-zero matrix elements of $\Omega_{n/\delta}$ can be written as:
\begin{equation}
\left( \Omega_{n/\delta} \right)_{\lambda\lambda'} = \sum_{\xi=0}^{\alpha-1} \delta_{\lambda', \,-\omega\lambda + 2\xi n/\alpha} \quad (\alpha | \lambda \mbox{ and } \alpha | \lambda')
\end{equation}
Changing variables by $\xi = \alpha - \xi'$ we have that
\begin{equation}
\left( \Omega_{n/\delta} \right)_{\lambda\lambda'} = \sum_{\xi'=1}^{\alpha} \delta_{\lambda', \,-\omega\lambda - 2\xi' n/\alpha + 2n} = \sum_{\xi'=0}^{\alpha-1} \delta_{\lambda', \,-\omega\lambda - 2\xi' n/\alpha} \quad (\alpha | \lambda \mbox{ and } \alpha | \lambda')
\end{equation}
So, when we compute the partition function, we find:
\begin{equation}
\label{4125}
Z_{n/\delta} = \frac{1}{2} \sum_{\lambda=1}^{2n} \sum_{\xi'=0}^{\alpha-1} \bar{\chi}_{\lambda}\,\chi_{-\omega\lambda - 2\xi' n/\alpha} = -\frac{1}{2} \sum_{\lambda=1}^{2n} \sum_{\xi'=0}^{\alpha-1} \bar{\chi}_{\lambda}\,\chi_{\omega\lambda + 2\xi' n/\alpha}= -Z_{\delta}
\end{equation}
What we saw above for the case of $\delta=1,n$ was a special case of this linear relation. 

To write a formula for the number of invariants, define the divisor function:
\be
\sigma(n)=\sum_{\delta |n}1
\ee
which counts the number of positive integers that divide $n$. Now if $\sigma(n)$ is even then each divisor $\delta$ has a distinct counterpart $\frac{n}{\delta}$ and in this case there are $\frac{\sigma(n)}{2}$ invariants. However if $\sigma(n)$ is odd then $\delta$ and $\frac{n}{\delta}$ must be the same for one value of $\delta$, say $\delta=m$. This means that $n=m^{2}$.  Using the relation (\ref{4125}), we have that $Z_{m} = - Z_{m}$, so $Z_{m}=0$. Therefore, for odd values of $\sigma(n)$, we have only $(\sigma(n)-1)/2$ linearly independent partition functions.

To summarise, the number of independent modular invariants is given by:
\begin{equation}
\label{4126}
\mbox{No. of linearly independent } Z_{\delta} = \begin{cases} \frac{\sigma(n)}{2}, \hspace{0.8cm} \sigma(n) \mbox{ even} \\ \frac{\sigma(n)-1}{2}, \quad \sigma(n) \mbox{ odd} \\ \end{cases}
\end{equation}
We now consider those heights $n$ that give rise to $1,2$ or 3 modular invariants. In the latter two cases one must additionally ensure that there are no unphysical invariants. In these cases the result of the \poinsum is either a CFT or an average over CFT's. Once there are four or more invariants, at least one is necessarily unphysical and the averaging interpretation no longer works. 

\subsubsection*{One Physical Invariant}

First let us ask when there is a unique modular invariant. From (\ref{4126}), we know that this happens when $\sigma(n)=2$, which means $n=p$ for some prime $p$, or $\sigma(n)=3$ in which case $n=p^2$ for some prime $p$. In these cases the unique modular invariant function is the one corresponding to $\Omega_{n}$ (defined in \eref{Omegasutwok}) and it is the diagonal invariant $Z_{A}$. When this is the case, the Poincar\'{e} sums that we compute must necessarily be proportional to $Z_A$, regardless of the seed. The central charges for these cases are of the form:
\begin{equation}
\bc = 3 - \frac{6}{p} \quad \hbox{or} \quad \bc = 3 - \frac{6}{p^{2	}}
\end{equation}
for some prime $p$. In Table \ref{ppsq} we list the first few examples of models which fall under this category. The models are labelled by the height $n=k+2$. 

\begin{longtable}{|>{\centering}p{2.7cm}|>{\centering}p{4cm}|}
\hline

 \hline
 \textbf{Level} $(n, \bc, N_{0})$  & \textbf{Poincar\'{e} sum}  \tabularnewline
 \hline
 \hline

$(3, 1, 12)$ & $24\, Z_{A}$  \tabularnewline
\hline

$(4, \frac{3}{2}, 16)$ & $32\, Z_{A}$ \tabularnewline
\hline

$(5, \frac{9}{5}, 20)$ & $36\, Z_{A}$ \tabularnewline
\hline

$(7, \frac{15}{7}, 28)$ & $24\, Z_{A}$ \tabularnewline
\hline

$(9, \frac{7}{3}, 36)$  & $54\, Z_{A}$ \tabularnewline
\hline


$(11, \frac{27}{11}, 44)$ & $54\, Z_{A}$ \tabularnewline
\hline

$(13, \frac{33}{13}, 52)$  & $84\, Z_{A}$ \tabularnewline
\hline

$(17, \frac{45}{17}, 68)$  & $108\, Z_{A}$ \tabularnewline

\hline

$(19, \frac{51}{19}, 76)$  & $120\, Z_{A}$ \tabularnewline

\hline

$(23, \frac{63}{23}, 92)$   & $144\, Z_{A}$ \tabularnewline

\hline

$(25, \frac{69}{25}, 100)$  & $150\, Z_{A}$ \tabularnewline

\hline

$(29, \frac{81}{29}, 116)$ & $180\, Z_{A}$ \tabularnewline
\hline

$(31, \frac{87}{31}, 124)$  & $192\, Z_{A}$ \tabularnewline
\hline

$(37, \frac{105}{37}, 148)$  & $228\, Z_{A}$ \tabularnewline
\hline


$(41, \frac{117}{41}, 164)$  & $252\, Z_{A}$ \tabularnewline
\hline

$(43, \frac{123}{43}, 172)$ & $264\, Z_{A}$ \tabularnewline
\hline

\caption{Poincar\'{e} sums for $SU(2)_{k}$ WZW models with unique modular invariant}
\label{ppsq}
\end{longtable}

Thus, these \poin sums  are candidate partition functions for a gravitational theory dual to the diagonal \sutk theory. It is intriguing that, unlike the Virasoro minimal models, here we find an infinite family of unitary models that pass the first test to be dual to a gravity theory.

\subsubsection*{Two Physical Invariants}

Next we consider the case when there are exactly two modular invariants. 
From (\ref{4126}) we see that this is the case when $\sigma(n)=4$ or $5$. For the former case, we have $n=pq$ or $n=p^3$ where $p,q$ are primes, while for the latter we have $n=p^4$. Since the diagonal invariant $Z_A$ always exists, this tells us there is exactly one more invariant for all these values of $n$. However, the second invariant is not necessarily physical. In fact, \cite{Cappelli:1986hf, Cappelli:1987xt} tells us that whenever there are two {\em physical} invariants, they must be as follows:
\be
Z_A=\bchi\,\Omega_n\chi,\qquad Z_D=\bchi(\Omega_2+\Omega_n)\chi
\label{ADinv}
\ee
Therefore we must consider only those sub-cases of $n=pq,p^3,p^4$ for which 2 divides $n$ (so that $\Omega_2$ can exist). These consist of one infinite family with $n=2p$ and two sporadic cases with $n=8,16$. Clearly these cases have no unphysical invariants. In Table \ref{2p} we list several examples of models which fall into this category, displaying the result of the \poinsum for all possible diagonal seeds of the form:
\be
(X_{\rm seed})_{\lambda,\lambda'}=\delta_{\lambda\lambda_0} \delta_{\lambda'\lambda_0}
\ee
(there can also be non-diagonal seeds corresponding to primaries with spin, these will come up later). The identity seed corresponds to $\lambda_0=1$. In the Table, the seed will be labelled by its value of $\lambda_0$.

\begin{longtable}{|>{\centering}p{2.5cm}|>{\centering}p{4cm}|>{\centering}p{4cm}|}
\hline

 \hline
 \textbf{Data} $(n, \bc, N_{0})$ & Value of $\lambda_0$  & \textbf{Poincar\'{e} sum} \tabularnewline
 \hline
 \hline


$(6, 2, 20)$ & $1,5$  & $8\, (4Z_{A} + Z_{D})$ \tabularnewline
		   & $2,4$ &   $32\, (2Z_{A} - Z_{D})$ \tabularnewline
		   & $3$ &  $48\, Z_{D}$ \tabularnewline
\hline

$(10, \frac{12}{5}, 40)$ & $1,3,7,9$ & $24\, (2 Z_{A} + Z_{D})$ \tabularnewline
   				     & $2,4,6,8$ & $48\, (2 Z_{A} - Z_{D})$ \tabularnewline
   				     & $5$ &  $96\, Z_{D}$ \tabularnewline
\hline

$(14, \frac{18}{7}, 56)$ & $1,3,5,9,11,13$ & $8\, (8Z_{A} + 5Z_{D})$ \tabularnewline
					  & $2,4,6,8,10,12$ &  $64\, (2Z_{A} - Z_{D})$ \tabularnewline
					  & $7$ &  $144\, Z_{D}$ \tabularnewline
\hline

$(22, \frac{30}{11}, 88)$ & ${\rm odd},\ne 11$ &  $24\, (4Z_{A} + 3Z_{D})$ \tabularnewline
					   & {\rm even} &   $96\, (2Z_{A} - Z_{D})$ \tabularnewline
					   & $11$ &   $240\, Z_{D}$ \tabularnewline

\hline

$(26, \frac{36	}{13}, 104)$ & ${\rm odd},\ne 13$ &  $8\, (14Z_{A} + 11Z_{D})$ \tabularnewline
						& {\rm even} &   $112\, (2Z_{A} - Z_{D})$ \tabularnewline
						& $13$ &   $288\, Z_{D}$ \tabularnewline
\hline

$(34, \frac{48}{17}, 136)$ & ${\rm odd},\ne 17$ & $24\, (6Z_{A} + 5Z_{D}) $ \tabularnewline
						& {\rm even} &  $144\, (2Z_{A} - Z_{D}) $ \tabularnewline
						& $17$ &   $384\, Z_{D} $ \tabularnewline
\hline


$(38, \frac{54}{19}, 152)$ & ${\rm odd},\ne 19$ & $ 8\, (20Z_{A} + 17Z_{D})$ \tabularnewline
						& {\rm even} &  $ 160\, (2Z_{A} - Z_{D})$ \tabularnewline
						& $19$ &   $ 432\, Z_{D}$ \tabularnewline
\hline

\caption{Poincar\'{e} Sums for $SU(2)_{k}$ WZW model with two modular invariants}
\label{2p}
\end{longtable}

We see in Table \ref{2p} that the linear combinations appearing for the case when the vacuum is the seed primary have positive coefficients for $Z_A$ and $Z_D$. From the perspective of AdS/CFT, this means that the gravity partition function could plausibly be interpreted as an average of the partition functions of an ensemble of physical CFTs. Here the ensemble has just 2 elements, a consequence of our choice of symmetry algebra. If we write the Poincare sum as $c_{A}Z_{A} + c_{D}Z_{D}$, then the relative size of the coefficients $c_{A}$ and $c_{D}$, provides the measure for how to average over the two theories. 

Following the discussion in the previous section, we now derive the coefficients $c_A,c_D$, namely the linear combinations of $Z_{A}$ and $Z_{D}$ that appear in the last column of Table \ref{2p} for any seed. This gives a successful prediction of all the results in Table \ref{2p} and extends them to $n=2p$ for arbitrary prime $p$. 

Referring to \eref{msolve}, we see that the matrix $d_{JK}$ is a $2 \times 2$ matrix in the present case, and is readily inverted. First, we need to find the relevant matrices $I_{J}$. Recall from the discussion around \eref{ADinv} that $\Omega_{n}$ and $\Omega_{2}$ exist. Let us now look at the modular invariant functions that we get from these matrices. We start with $\Omega_{n}$, so $\alpha=1$ and $\omega=1$ and the invariant is:
\begin{equation}
Z_{n} = \frac{1}{2}\sum_{\lambda,\lambda'=1}^{2n} \bar{\chi}_{\lambda} (\Omega_{n})_{\lambda\lambda'} \chi_{\lambda'} 
= \frac{1}{2}\sum_{\lambda,\lambda'=1}^{2n} \bar{\chi}_{\lambda} \delta_{\lambda\lambda'} \chi_{\lambda'} = \frac{1}{2}\sum_{\lambda=1}^{2n} \abs{\chi_{\lambda}}^{2} 
= \sum_{\lambda=1}^{n-1} \abs{\chi_{\lambda}}^{2}  
\end{equation}
where we have used $\chi_{\lambda} = -\chi_{-\lambda} = \chi_{\lambda+2n}$. Thus we see that $\Omega_{n}$ when acting on the restricted space ($1\le \lambda\le n-1$) still operates as the identity matrix. Hence $(I_{n})_{\lambda\lambda'} = \delta_{\lambda\lambda'}$, where $\lambda,\lambda'$ takes values from $1$ to $n-1$. Next we consider $\Omega_{2}$. Here $\alpha=[2,p]=1$, since $n=2p$ and $p>2$. We need to chose $\omega^{2}=1$ mod $8p$. For this, we can chose $\omega=2p-1=n-1$ which is allowed since $\omega^{2} = 4(p-1)p + 1=1$ mod $8p$. The associated modular invariant is:
\begin{equation}
\begin{split}
Z_{2} &= \frac{1}{2}\sum_{\lambda,\lambda'=1}^{2n} \bar{\chi}_{\lambda} (\Omega_{2})_{\lambda\lambda'} \chi_{\lambda'} = \frac{1}{2}\sum_{\lambda,\lambda'=1}^{2n} \bar{\chi}_{\lambda} \, \delta_{\lambda',(n-1)\lambda} \, \chi_{\lambda'} \\
&= \frac{1}{2}\sum_{\lambda=1}^{2n} \bar{\chi}_{\lambda} \, \chi_{(n-1)\lambda} 
= \frac{1}{2}\sum_{\lambda \, {\rm even} }^{2n} \bar{\chi}_{\lambda} \, \chi_{(n-1)\lambda} + \frac{1}{2}\sum_{\lambda \, {\rm odd}}^{2n} \bar{\chi}_{\lambda} \, \chi_{(n-1)\lambda} 
\end{split}
\end{equation}
Now, if $\lambda$ is even then in the first sum we can shift the subscript $\chi_{(n-1)\lambda}$ by some multiple of $2n$ and bring it to $\chi_{-\lambda}=-\chi_{\lambda}$. Similarly, for the case of odd $\lambda$, we can shift the subscript to bring it to $\chi_{n-\lambda}$. Therefore we have:
\begin{equation}
\begin{split}
Z_{2} &= -\frac{1}{2}\sum_{\lambda \, {\rm even}}^{2n} \abs{\chi_{\lambda}}^{2} + \frac{1}{2}\sum_{\lambda \, {\rm odd}}^{2n} \bar{\chi}_{\lambda} \, \chi_{n-\lambda} \\
&= -\sum_{\lambda \, {\rm even}}^{n-1} \abs{\chi_{\lambda}}^{2} + \sum_{\lambda {\rm odd}}^{n-1} \bar{\chi}_{\lambda} \, \chi_{n-\lambda} \\
\end{split}
\end{equation}
So the matrix elements for $I_{2}$ can be written:
\begin{equation}
(I_{2})_{\lambda\lambda'} = \begin{cases} -\delta_{\lambda\lambda'}~, \quad \lambda \mbox{ even} \\ \delta_{\lambda',n-\lambda}~, \hspace{0.35cm}\lambda \mbox{ odd} \end{cases}
\end{equation}
Note that there are $\frac{n-1}{2} = p-1$ even values and $\frac{n}{2} = p$ odd values for $\lambda$. Also, the only diagonal term for odd $\lambda$ occurs at $\lambda = \frac{n}{2} = p$. 

With this information we are now equipped to compute the matrix $d_{JK}$ of inner products defined in \eref{dmat}, where $J,K$ take the two values $2,n$. We have:\begin{equation}
\begin{split}
\mbox{Tr}(I_{n}^{2}) &=  2p-1\\
\mbox{Tr}(I_{n}\,I_{2}) &=  2-p\\
\mbox{Tr}(I_{2}^{2}) &= 2p-1
\end{split}
\label{innprod}
\end{equation}
Now \eref{dmateqn} becomes:
\begin{equation}
\begin{pmatrix} 2p-1 & 2-p\\ 2-p & 2p-1  \end{pmatrix}
\begin{pmatrix} c_{n}\\ c_{2} \end{pmatrix}
=
\abs{\Gamma_{\rm sub}\backslash\Gamma}
\begin{pmatrix} 
\mbox{Tr}(I_{n}X_{{\rm seed}}) \\ \mbox{Tr}(I_{2}X_{\rm seed}) \end{pmatrix}
\label{mateqn}
\end{equation}
where we have taken $\sum_{\gamma}M^{\dagger}_{\gamma}X_{\rm seed}M_{\gamma} = c_{n}I_{n} + c_{2}I_{2}$. Solving the above equation, we get:
\begin{equation}
\begin{split}
\begin{pmatrix} c_{n}\\ c_{2} \end{pmatrix}
&= \frac{\abs{\Gamma_{\rm sub}\backslash\Gamma}}{3(p^{2}-1)}
\begin{pmatrix} 2p-1 & p-2 \\ p-2 & 2p-1  \end{pmatrix}
\begin{pmatrix} \Tr(I_{n}X_{\rm seed}) \\ \mbox{Tr}(I_{2}X_{\rm seed}) \end{pmatrix} \\
&= \frac{\abs{\Gamma_{\rm sub}\backslash\Gamma}}{3(p^{2}-1)}
\begin{pmatrix} (2p-1)\mbox{Tr}(I_{n}X_{\rm seed}) + (p-2)\mbox{Tr}(I_{2}X_{\rm seed}) \\ (p-2)\mbox{Tr}(I_{n}X_{\rm seed}) + (2p-1)\mbox{Tr}(I_{2}X_{\rm seed})  \end{pmatrix}\\
\end{split}
\end{equation}
This is the general expression for the result of a Poincar\'e sum for an arbitrary seed at level $k=n-2=2p-2$. 

Let us now look at all possible individual cases of $X_{\rm seed}$ and see how they explain the data that we found in Table \ref{2p}. We will use the fact that $Z_{n}=Z_{A}$ and $Z_{2} = Z_{D} - Z_{A}$, from which it follows that $c_A=c_n-c_2, c_D=c_2$. 

\begin{itemize}
\item $(X_{\rm seed})_{\lambda\lambda'} = \delta_{\lambda1}\delta_{\lambda'1}$. In this case, $\mbox{Tr}(I_{n}X_{\rm seed}) = 1$ and $\mbox{Tr}(I_{2}X_{\rm seed}) = 0$. This means that $c_{n}=2p-1$ and $c_{2}=p-2$. So the Poincar\'e sum is proportional to:
\begin{equation}
(2p-1)Z_{n} + (p-2)Z_{2} = (p+1)Z_{A} + (p-2)Z_{D}
\label{idavg}
\end{equation}
The ratio of the two coefficients is $\frac{c_{D}}{c_{A}} = \frac{p-2}{p+1} = \frac{n-4}{n+2} = \frac{k-2}{k+4}$. In the limit $p\rightarrow{\infty}$, we have $\frac{c_{D}}{c_{A}} \rightarrow{1}$. One can also verify that all diagonal seeds of the form $(X_{\rm seed})_{\lambda,\lambda'} = \delta_{\lambda,\lambda_0}\delta_{\lambda',\lambda_0}$, where $\lambda_0$ is odd but not equal to $p$, give the same result.

\item $(X_{\rm seed})_{\lambda\lambda'} = \delta_{\lambda2}\delta_{\lambda'2}$. This gives $\mbox{Tr}(I_{n}X_{\rm seed}) = 1$ and $\mbox{Tr}(I_{2}X_{\rm seed}) = -1$ and hence $c_{n}=p+1$ and $c_{2}=-(p+1)$. So the \poinsum is proportional to:
\be
(p+1)Z_n - (p+1)Z_2=2(p+1)Z_A-(p+1)Z_D
\ee
One can also verify that all diagonal seeds of the form $(X_{\rm seed})_{\lambda,\lambda'} = \delta_{\lambda,\lambda_0}\delta_{\lambda',\lambda_0}$, where $\lambda_0$ is even, give the same result.

\item $(X_{\rm seed})_{\lambda\lambda'} = \delta_{\lambda p}\delta_{\lambda' p}$. This gives $\mbox{Tr}(I_{n}X_{\rm seed}) = 1$ and $\mbox{Tr}(I_{2}X_{\rm seed}) = 1$ and hence $c_{n}=3(p-1)$ and $c_{2}=3(p-1)$. In this case  the Poincar\'e sum is proportional to:
\begin{equation}
(3p-3)Z_{n} + (3p-3)Z_{2} = (3p-3)Z_{D}
\end{equation}
Thus for models with $k+2=2p$, if we start with the seed $\chi_{p}$ then the Poincar\'e sum gives us only the $Z_{D}$ invariant. 

\item $(X_{\rm seed})_{\lambda\lambda'} = \delta_{\lambda1}\delta_{\lambda',2p-1} \implies \mbox{Tr}(I_{n}X_{\rm seed}) = 0$ and $\mbox{Tr}(I_{2}X_{\rm seed}) = 1$. This means that $c_{n}=p+1$ and $c_{2}=-(p+1)$. So, the Poincar\'e sum is 
\begin{equation}
\label{4145}
P_{1,2p-1} \propto (p-2)Z_{n} + (2p-1)Z_{2} = -(p+1)Z_{A} + (2p-1)Z_{D}
\end{equation}
this expression holds true for any odd $\lambda \,(\ne p)$ because $\mbox{Tr}(I_{n}X_{\rm seed}) = 0$ and $\mbox{Tr}(I_{2}X_{\rm seed}) = 1$ still holds true.
\end{itemize}

One can now verify that the above expressions exactly match the linear combinations appearing in Table \ref{2p}. In Table \ref{2pgen} we summarise the above results by listing the values of $c_A,c_D$ in each category.

\begin{table}[h!]
  \centering
  \begin{tabular}{|c|c|}
    \hline
    Seed &  $c_A,c_D$  \\
    \hline
$\lambda_0=\lambda_0'={\rm odd} \ne p$  & $p+1,p-2$\\
$\lambda_0=\lambda_0'=$ even  & $2(p+1),-(p+1)$\\
 $\lambda_0=\lambda_0'=p$  & $(0,3p-3)$\\
 $\lambda_0={\rm odd}\ne p,\lambda_0'=2p-\lambda_0$ &
 $-(p+1),2p-1$\\
 \hline
  \end{tabular}
  \caption{General result, two modular invariants. The seeds are labelled by $\lambda_0,\lambda_0'$ where $(X_{\rm seed})_{\lambda\lambda'}=\delta_{\lambda\lambda_0}\delta_{\lambda'\lambda_0'}$}
\label{2pgen}
\end{table}

So far we have worked with individual (and diagonal) seeds. More generally we can consider linear combinations of different seeds. For future use, we would like to find the seed that will give us an arbitrary general linear combination $\alpha Z_A +\beta Z_D$. Since there are only two terms in the combination but more than two possible seeds, this can clearly always be done. Let us therefore find the most general seed $Z_{\rm seed}(\alpha,\beta)$ that satisfies:
\be
\sum_\gamma Z_{\rm seed}(\alpha,\beta)=\alpha Z_A+\beta Z_D
\ee
The general solution (now also including possible non-diagonal seeds) is:
\begin{equation}
Z_{\rm seed}(\alpha,\beta) =a_{1} \!\!\!\sum_{\substack{\lambda\, {\rm odd},\,  \ne p}}\!\!\! b_{\lambda}\abs{\chi_{\lambda}}^{2}  + a_{2}\!\! \sum_{\substack{\lambda\, \rm even}}\!\!c_{\lambda}\abs{\chi_{\lambda}}^{2}  + a_{3}\abs{\chi_{p}}^{2}
 + a_{4} \sum_{\lambda\,{\rm odd},=1}^{p-2}d_{\lambda}(\bchi_{\lambda}\chi_{2p-\lambda} + {\rm c.c.}) 
 \label{4189}
\end{equation}
where:
\begin{equation}
\label{4190}
\begin{split}
&\hspace{2cm}\sum_{\lambda} b_{\lambda} = \sum_{\lambda} c_{\lambda} = \sum_{\lambda} d_{\lambda} = 1 \\
&a_{3} = \frac{(2p-1)\alpha+(p+1)\beta}{3(p^{2}-1)} - (a_{1}+a_{2}), \quad 2a_{4} = a_{1} + 2a_{2} - \frac{\alpha}{p+1}
\end{split}
\end{equation}
It is easily verified that the Poincar\'e sum over (\ref{4189}) gives $\alpha Z_A+\beta Z_D$ as desired. Also, special values of the coefficients $a_i,b_\lambda$ reproduce the single-seed results: for example, choosing $a_1=1, a_2=a_3=a_4=0$ and $b_{\lambda_0}=1$ for some odd $\lambda_0$  reproduces the answer $(p+1)Z_A+(p-2)Z_D$, and similarly for the other cases. 

\subsubsection*{Three Physical Invariants}

Finally we turn to the cases with three physical  invariants, corresponding to $\sigma(n)=6, 7$. Here a fresh analysis is not necessary. From \cite{Cappelli:1986hf, Cappelli:1987xt} we find that there are precisely three invariants, all physical, only for $k=10,16$. Let us write $Z_{A/D/E}=\bchi \,\Omega_{A/D/E}\chi$. Then, in the first case we have the matrices $\Omega_2,\Omega_3,\Omega_{12}$ from which we get $\Omega_A=\Omega_{12}, \Omega_D=\Omega_{12}+\Omega_2$ and $\Omega_E=\Omega_{12}+\Omega_3+\Omega_2$, the last one corresponding to the $E_6$ invariant. In the second case the matrices are $\Omega_2,\Omega_3, \Omega_{18}$ from which one gets $\Omega_A=\Omega_{18}, \Omega_D=\Omega_{18}+\Omega_2$ and $\Omega_E=\Omega_{18}+\Omega_3+\Omega_2$, of which the last one corresponds to the $E_7$ invariant. For $k=28$ we instead get {\em four} matrices $\Omega_2,\Omega_3, \Omega_5, \Omega_{30}$ but the physical partition functions comes from just three matrices $\Omega_A=\Omega_{30}, \Omega_D=\Omega_{30}+\Omega_2$ and $\Omega_E=\Omega_{30}+\Omega_5+\Omega_3+\Omega_2$ with the last one being associated to the $E_8$ invariant. However, since there are four invariants in all but only three physical invariants, there is one independent unphysical linear combination that can appear in the \poinsum. In Table \ref{SU2fulltable} we see that this unphysical invariant does make an appearance in the \poinsum, so this case cannot be interpreted as an average over three RCFT's. 

\subsection{SU($N$)$_k$ WZW Models}

We now move on to discuss the case of \sunk for $N >2$. We will encounter both similarities and differences with respect to the \sutk case analysed above. One of the key differences is that all representations of SU(2) are real or pseudo-real but for SU($N$), $N>3$ there are both complex and real representations. For each complex representation, the CFT will contain two primaries (one being the complex conjugate of the other) that share the same character. This results in a multiplicity of 2 for such representations. If ignored, this fact can lead to incorrect results, so our strategy will be to treat a representation and its complex conjugate as distinct at intermediate stages of the computations, and then identify them at the end.

Using the data in Appendix \ref{basicdata} one can in principle compute the Poincar\'e sums to obtain candidate gravity partition functions for any \sunk, using the general procedure described earlier. The idea, as before, is to identify models that have only physical modular invariants, then express the Poincar\'e sums in terms of these invariants and attempt to interpret the linear combination as an average over CFT's. 

While modular invariants for SU(2)$_{k}$ were completely classified in \cite{Cappelli:1986hf, Cappelli:1987xt}, the case of SU(3)$_{k}$ is considerably more complicated and was carried out later in \cite{Gannon:1992ty, Gannon:1994cf}. The classification of modular invariants for the SU($N$)$_{1}$ series was conjectured in \cite{Itzykson:1988hk} and proved in \cite{Degiovanni:1989ne}. Some results for the general case of \sunk can been found in \cite{Bauer:1990xx}. In what follows, we will focus on the families \sunone and \suthk, for which the general classification is well-understood and reasonably tractable. 

\subsubsection{Poincar\'e Sums for SU($N$)$_1$}

The case of \sunone shares several features with \sutk WZW models. In particular, we will see below that the modular invariants can be written down in a similar way as for SU(2)$_{k}$. In \sunone there are $N$ allowed representations including the identity. Using (\ref{673}), we see that the allowed Dykin labels are given by $\lambda_{i}=\rho_{i}=1$ or $\lambda_{i}=\rho_{i} + \delta_{b,i}=1+\delta_{b,i}$ for some $b$ between $1$ to $N-1$. So, we can denote these as $\lambda^{a}$ where $a\in\{0,1,\dots,N-1\}$ and $(\lambda^{a})_{i} = 1 + \delta_{a,i}$. The characters can now be labelled by the index $a$. 

The central charge and the conformal dimensions are:
\begin{align}
\mathbf{c} &= N-1 \\
h_{a} &= \frac{a(N-a)}{2N} 
\end{align} 
For $a>0$ the representations labelled by $a$ and $N-a$ are complex conjugates of each other, due to which the characters labelled by $a$ and $N-a$ are identical. The fact that $h_{a}=h_{N-a}$ is consistent with this identification. This leads to the multiplicity 2 in the partition function that was referred to above. 

Following \cite{Itzykson:1988hk, Degiovanni:1989ne}, we now list the modular invariants that appear in these theories. We define the integer $m$ by:
\begin{equation}
m = \begin{cases} N  \mbox{, if $N$ is odd} \\  \frac{N}{2}  \mbox{, if $N$ is even} \end{cases}
\label{mdef}
\end{equation}
This integer plays a somewhat similar (but not identical) role to the height $n=k+2$ for \sutk theories. 
Now for every divisor $\delta$ of $m$, define $\alpha\equiv[\delta,m/\delta]$ (the GCD of the integers) and let $\omega(\delta) = \left(\rho\frac{m}{\alpha\delta}+\sigma\frac{\delta}{\alpha}\right) \mbox{ mod } \frac{N}{\alpha^{2}}$, where we have chosen integers $\rho,\sigma$ such that  $\rho\frac{m}{\alpha\delta}-\sigma\frac{\delta}{\alpha}=1$. This is always possible because $\frac{m}{\alpha\delta}$ and $\frac{\delta}{\alpha}$ are coprime. We note that $\omega(m/\delta)=-\omega(\delta) \mbox{ mod }  \frac{N}{\alpha^{2}}$. Finally, we define the matrices $\Omega_{\delta}$:
\begin{equation}
\left( \Omega_{\delta} \right)_{aa'} = \begin{cases} 0, \hspace{3.59cm} \alpha \nmid a \mbox{ or } \alpha \nmid a'    \\ \sum_{\xi=0}^{\alpha-1} \delta_{a', \, \omega(\delta)a + \xi N/\alpha}, \quad \mbox{otherwise} \end{cases}
\label{Omegasunone}
\end{equation}
The modular invariant partition functions are then:
\begin{equation}
\label{683}
Z_{\delta} = \sum_{a=0}^{N-1}\bar{\chi}_{a} \left( \Omega_{\delta} \right)_{aa'} \chi_{a'}
\end{equation}
Here the indices $a,a'$ in the matrix elements and characters will always be understood as integers modulo $N$. Not all the $Z_{\delta}$ above are independent, in fact  $Z_{\delta}$ and $Z_{m/\delta}$ are equal. This can be seen from the fact that:
\begin{equation}
\begin{split}
(\Omega_{m/\delta})_{aa'} = \sum_{\xi=0}^{\alpha-1} \delta_{a', \, -\omega(\delta)a + \xi N/\alpha} = \sum_{\xi'=0}^{\alpha-1} \delta_{a', \, N - \omega(\delta)a - \xi' N/\alpha}  \quad (\alpha|a \mbox{ and } \alpha|a') \\
\end{split}
\end{equation}
and the corresponding partition function is:
\begin{equation}
\begin{split}
Z_{m/\delta} &= \sum_{\substack{a=0 \\ \alpha | a}}^{N-1} \sum_{\xi'=0}^{\alpha-1} \bar{\chi}_{a}\,  \chi_{N -\omega(\delta)a - \xi' N/\alpha} = \sum_{\substack{a=0 \\ \alpha | a}}^{N-1} \sum_{\xi'=0}^{\alpha-1} \bar{\chi}_{a}\,  \chi_{\omega(\delta)a + \xi' N/\alpha}= Z_{\delta} \\
\end{split}
\end{equation}
In the above, we used $\xi'=\alpha-\xi$ and $\chi_{b} = \chi_{N-b}$. Thus we have:
\begin{equation}
\label{686}
\mbox{Number of linearly independent } Z_{\delta} = \begin{cases} \frac{\sigma(m)}{2}, \hspace{0.8cm} \sigma(m) \mbox{ is even} \\ \frac{\sigma(m)+1}{2}, \quad \sigma(m) \mbox{ is odd} \\ \end{cases}
\end{equation}
Remarkably, for \sunone all the $Z_{\delta}$ are not just modular invariant but also physical -- they satisfy positive integrality of coefficients and non-degeneracy of the vacuum state. This is a major difference from \sutk. Another difference is that while the latter can have at most three physical invariants namely $Z_A, Z_D, Z_E$, there is no such restriction for \sunone where one can have an arbitrarily large number of physical modular invariants for sufficiently large and suitably chosen $N$. 

We now perform a similar classification as before to find conditions for one invariant, two invariants and so on. We also derive the linear combinations of the invariants that will appear by inverting the matrix of inner products among the modular invariant basis matrices. To do this, we will use all the available $\Omega_\delta$ as the basis for a given model and then compute the cofficients $c_\delta$ corresponding to the modular invariant $Z_\delta$. At the end, we will impose the constraint that $Z_{m/\delta}=Z_\delta$ which equates the label for a representation and its complex conjugate. 

\subsubsection*{One Physical Invariant}

These models correspond to what we have called type (I). As can be seen from (\ref{686}), we have a unique invariant whenever $\sigma(m)=1$ or $2$. This is true when $m=1$ or $p$, where $p$ is a prime. So altogether we have $N=2,p,2p$ where $p$ is prime. In Table \ref{sunone.1} we list some examples of models of this type.

\begin{longtable}{|>{\centering}p{3cm}|>{\centering}p{4cm}|>{\centering}p{4cm}|}
\hline

 \hline
 \sunone\\ $ (N, c, N_{0})$ & \textbf{Seed Primary} $(\chi_{\lambda})$ & \textbf{Poincar\'{e} sum}   \tabularnewline
 \hline
 \hline

$(2, 1, 4)$ & $0,1$ & $3\, Z_p $  \tabularnewline
\hline

$(3,2,6)$ & $0$ & $6\, Z_p $ \tabularnewline
                & $1,2$ & $3\, Z_p $ \tabularnewline
\hline

$(4, 3,8)$ & $0,2$  & $8\, Z_p $ \tabularnewline
				 & $1,3$  & $4\, Z_p $ \tabularnewline
\hline

$(5,4,10)$ & $0$  & $12\, Z_p $ \tabularnewline
				 & $1,2,3,4$  & $6\, Z_p $ \tabularnewline
\hline

$(6,5,12)$ & $0,3$  & $12\, Z_p $ \tabularnewline
				 & $1,2,4,5$  & $6\, Z_p $ \tabularnewline
\hline

$(7,6,14)$ & $0$  & $18\, Z_p $ \tabularnewline
				 & $1,2,3,4,5,6$  & $9\, Z_p $ \tabularnewline
\hline

$(10,9,20)$ & $0,5$  & $24\, Z_p $ \tabularnewline
				 & $1,2,3,4,6,7,8,9$  & $12\, Z_p $ \tabularnewline
\hline


$(19,18,38)$ & $0$  & $54\, Z_p $ \tabularnewline
				 & $1,2,\cdots,18$  & $27\, Z_p $ \tabularnewline
\hline

\caption{Poincar\'{e} sums for \sunone WZW models with a unique modular invariant.}
\label{sunone.1}
\end{longtable}
Just like the SU(2)$_{k}$ examples with unique modular invariants listed in Table \ref{ppsq}, the cases in Table \ref{sunone.1} provide candidate gravity partition functions dual to a unique \sunone WZW model.

\subsubsection*{Two physical invariants}

Now, we consider the case where there are exactly two modular invariants. From (\ref{686}), we see that here $\sigma(m)=3$ or $4$ which is true when $m=p^{2},\, p^{3}, \, pq$, where $p,q$ are primes and $m>5$. So the relevant values of $N$ are $N=p^{2},\,p^{3}, pq,   2p^{2},\,2p^{3}$ or $2pq$.

All these values of $N$ describe models which fall into the type (IIa) category. Unlike \sutk, we do not need to restrict further to sub-families as all of them are physical. In Table \ref{sunone.2} we list examples of models having $m=p^2$, i.e. $N=p^2$ or $2p^2$, along with the result of the Poincar\'e sum for each seed primary. In these models the possible values of the integers $\delta$ that divide $m$ are $1,p,p^2$ and correspondingly there are three modular invariants that we label $Z_1,Z_p,Z_{p^2}$. We have $Z_1=Z_{p^2}$ resulting from complex conjugation, as described above, so the results can only be linear combinations of $Z_p,Z_{p^2}$.

\begin{longtable}{|>{\centering}p{3cm}|>{\centering}p{4cm}|>{\centering}p{4cm}|}
\hline
 $SU(N)_{1}$\\ $ (N, c, N_{0})$ & \textbf{Seed Primary} $(\chi_{\lambda})$ & \textbf{Poincar\'{e} sum}   \tabularnewline
 \hline
 \hline

$(8, 7, 16)$ & $0,4$ & $ 4(4Z_{p^{2}} + Z_{p})$  \tabularnewline
					& $0 \ne \mbox{ mod } 2$ & $ 8(2Z_{p^{2}} - Z_{p})$  \tabularnewline
					& $0 = \mbox{ mod } 2,\, \ne0,4$ & $ 12Z_{p}$  \tabularnewline
\hline
$(9, 8, 18)$ & $0$ & $ 6(3Z_{p^{2}} + Z_{p})$  \tabularnewline
					& $0 \ne \mbox{ mod } 3$ & $ 4.5(3Z_{p^{2}} - Z_{p})$  \tabularnewline
					& $0 = \mbox{ mod } 3,\, \ne0$ & $ 1.5(3Z_{p^{2}} + 7Z_{p})$  \tabularnewline
\hline

$(18, 17, 36)$ & $0,9$ & $ 12(3Z_{p^{2}} + Z_{p})$  \tabularnewline
					& $0 \ne \mbox{ mod } 3$ & $ 9(3Z_{p^{2}} - Z_{p})$  \tabularnewline
					& $0 = \mbox{ mod } 3,\, \ne0,9$ & $ 3(3Z_{p^{2}} + 7Z_{p})$  \tabularnewline
\hline

$(25,24,50)$ & $0$ & $12(5Z_{p^{2}} + 2Z_{p})$ \tabularnewline
                     & $0 \ne \mbox{ mod } 5$ & $7.5(5Z_{p^{2}} - Z_{p})$ \tabularnewline
                     & $0 = \mbox{ mod } 5,\, \ne0$ & $4.5(5Z_{p^{2}} + 7 Z_{p})$ \tabularnewline
\hline

$(49,48,98)$ & $0$ & $18( 7Z_{p^{2}} + 3Z_{p})$ \tabularnewline
                     & $0 \ne \mbox{ mod } 7$ & $10.5( 7Z_{p^{2}} - Z_{p})$  \tabularnewline
                     & $0 = \mbox{ mod } 7,\, \ne0$ & $1.5( 35Z_{p^{2}} + 43Z_{p})$ \tabularnewline
\hline

$(50,49,100)$ & $0,25$ & $24(5Z_{p^{2}} + 2Z_{p})$ \tabularnewline
                     & $0 \ne \mbox{ mod } 5$ & $15(5Z_{p^{2}} - Z_{p})$ \tabularnewline
                     & $0 = \mbox{ mod } 5,\, \ne0,25$ & $9(5Z_{p^{2}} + 7 Z_{p})$ \tabularnewline
\hline

\caption{Poincar\'{e} sums for $SU(N)_{1}$ with two invariants, for $N=p^2,2p^2$}
\label{sunone.2}
\end{longtable}

These Poincar\'e sums provide candidate gravity partition functions where the dual consists of an average over the two CFTs. The precise linear combinations of the physical CFT partition function that we obtain can be derived using methods similar to those for $SU(2)_{k}$. We illustrate this for the same sub-families $N=p^{2}$ and $N=2p^2$, that were explicitly studied above. One can easily generalise this derivation to any of the other two-invariant families.

\subsubsection*{Two invariants: Coefficients for $N=p^2$}

Here, $m=N=p^{2}$ and from the discussion of the available modular invariant partition functions around (\ref{683}), the relevant matrices are $\Omega_{p^{2}},\, \Omega_{1},$ and $ \Omega_{p}$. Their matrix elements are given by:
\begin{equation}
(\Omega_{p^{2}})_{aa'} = \delta_{a'a} \,,\quad (\Omega_{1})_{aa'} = \delta_{a',-a} \,,\quad (\Omega_{p})_{aa'} = \begin{cases} 0, \hspace{2.1cm} p \nmid a \mbox{ or } p \nmid a'    \\ \sum_{\xi=0}^{p-1} \delta_{a',  \xi p}, \quad \mbox{otherwise} \end{cases}
\end{equation}
For the case of $\Omega_{p}$, we have used the fact that $\alpha\equiv[p,p]=p$ and so $N/\alpha^{2}=1$. Hence, $\omega(p)=0$.
The modular invariant functions that we get from these matrices are:
\be
\begin{split}
Z_{p^{2}} &= \sum_{a,a'=0}^{N-1}\bar{\chi}_{a} \left( \Omega_{p^{2}} \right)_{aa'} \chi_{a'} = \sum_{a,a'=0}^{N-1}\bar{\chi}_{a}\delta_{a'a}\chi_{a'} = \sum_{a=0}^{N-1}\abs{\chi_{a}}^{2} \\
Z_{1} &= \sum_{a,a'=0}^{N-1}\bar{\chi}_{a} \left( \Omega_{1} \right)_{aa'} \chi_{a'} = \sum_{a,a'=0}^{N-1}\bar{\chi}_{a}\delta_{a',-a}\chi_{a'} = \sum_{a=0}^{N-1}\bar{\chi}_{a}\chi_{-a} \\
Z_{p} &= \sum_{a,a'=0}^{N-1}\bar{\chi}_{a} \left( \Omega_{p} \right)_{aa'} \chi_{a'} = \sum_{\substack{a,a'=0 \\ p|a \, , \, p|a'}}^{N-1}\sum_{\xi=0}^{p-1} \bar{\chi}_{a} \,  \delta_{a',\xi p} \,  \chi_{a'}\\
&\qquad = \sum_{\substack{a=0 \\ p|a}}^{N-1} \bar{\chi}_{a} \sum_{\xi=0}^{p-1} \chi_{\xi p} = \left| \sum_{a=0,\, p|a}^{N-1}\chi_{a} \right|^{2}
\end{split}
\label{sunoneinv}
\ee
From the fact that $\chi_{-a}=\chi_{N-a}=\chi_a$, we see that $Z_1=Z_{p^2}$. However, as explained above, we treat them as independent for now and identify them at the end.

We see that the theories corresponding to $Z_{p}$ are holomorphically factorised, or meromorphic, CFT's. This can be understood as follows. The central charge of $SU(N)_{1}$ with $N = p^{2}$ is $p^{2}-1$. It is well-known that for any prime $p$, this number is divisible by $24$. Now $\mathbf{c}$ being a multiple of $24$ is a necessary condition to encounter meromorphic (one-character) CFT's, and these are what the $Z_{p}$ are giving us. As an example, the $Z_{p}$ invariant for $SU(25)_{1}$ is entry No. $67$ in the table of \cite{Schellekens:1992db}, in which all meromorphic CFT with $\mathbf{c}= 24$ are classified. 

We can now compute the matrix $d_{JK}$ of inner products. Its elements are:
\begin{align}
&\mbox{Tr}(\Omega_{p^{2}}^{2}) = p^{2} \,,\quad \mbox{Tr}(\Omega_{1}^{2}) = p^{2} \,,\quad \mbox{Tr}(\Omega_{p}^{2}) = p^{2} \\
\mbox{Tr}(\Omega_{p^{2}}\Omega_{1}) = &\mbox{Tr}(\Omega_{1}) = 1 \,,\quad \mbox{Tr}(\Omega_{p^{2}}\Omega_{p}) = \mbox{Tr}(\Omega_{p}) = p \,,\quad \mbox{Tr}(\Omega_{1}\Omega_{p}) = p
\end{align}
The linear combinations that appear in the Poincar\'e sums can now be computed:
\be
\begin{split}
\begin{pmatrix} c_{p^{2}}\\ c_{1} \\ c_{p} \end{pmatrix}
&= \abs{\Gamma_{\rm sub}\backslash \Gamma}
\begin{pmatrix} p^{2} & 1 & p \\ 1 & p^{2} & p \\ p & p & p^{2}   \end{pmatrix}^{-1}
\begin{pmatrix} \mbox{Tr}(\Omega_{p^{2}}X_{\rm seed}) \\ \mbox{Tr}(\Omega_{1}X_{\rm seed}) \\ \mbox{Tr}(\Omega_{p}X_{\rm seed})  \end{pmatrix} \\
&= \frac{\abs{\Gamma_{\rm sub}\backslash \Gamma}}{p^2(p^2-1)}\begin{pmatrix} p^{2} & 0 & -p \\ 0 & p^{2} & -p \\ -p & -p & 1+p^{2}  \end{pmatrix} \begin{pmatrix} \mbox{Tr}(\Omega_{p^{2}}X_{\rm seed}) \\ \mbox{Tr}(\Omega_{1}X_{\rm seed}) \\ \mbox{Tr}(\Omega_{p}X_{\rm seed})  \end{pmatrix} \\
&=\frac{\abs{\Gamma_{\rm sub}\backslash \Gamma}}{p^2(p^2-1)}
\begin{pmatrix} p^{2}\,\mbox{Tr}(\Omega_{p^{2}}X_{\rm seed}) - p\,\mbox{Tr}(\Omega_{p}X_{\rm seed})  \\ p^{2}\,\mbox{Tr}(\Omega_{1}X_{\rm seed}) - p\,\mbox{Tr}(\Omega_{p}X_{\rm seed})  \\   - p\,\mbox{Tr}(\Omega_{p^{2}}X_{\rm seed})  - p\,\mbox{Tr}(\Omega_{1}X_{\rm seed})  +(1+p^{2})\mbox{Tr}(\Omega_{p}X_{\rm seed})   \end{pmatrix}\\
\end{split}
\ee
Let us now consider different choices of seeds:
\begin{itemize}
\item $(X_{\rm seed})_{a,a'} = \delta_{a,0}\delta_{a',0}$. We have $\mbox{Tr}(\Omega_{p^{2}}X_{\rm seed}) = \mbox{Tr}(\Omega_{1}X_{\rm seed}) = \mbox{Tr}(\Omega_{p}X_{\rm seed}) = 1$ so the Poincar\'e sum is proportional to:
\begin{equation}
p(p-1)(Z_{p^{2}} + Z_{1}) + (p-1)^{2}Z_{p} = 2(p-1)\left( p\, Z_{p^{2}} + \frac{(p-1)}{2}Z_{p} \right)
\end{equation}
To get the second equality we have used $Z_{p^{2}}=Z_{1}$.
\item $(X_{\rm seed})_{a,a'} = \delta_{a,1}\delta_{a',1}$. We have $\mbox{Tr}(\Omega_{p^{2}}X_{\rm seed}) = 1\,, \mbox{Tr}(\Omega_{1}X_{\rm seed}) = 0 \,, \mbox{Tr}(\Omega_{p}X_{\rm seed}) = 0$ and so the Poincar\'e sum is proportional to:
\begin{equation}
p^{2}Z_{p^{2}} - p Z_{p} = p( p\,Z_{p^{2}} - Z_{p})
\end{equation}
This is true for any seed primary $\abs{\chi_{a}}^{2}$ where $p\nmid a$.
\item $(X_{\rm seed})_{a,a'} = \delta_{a,p}\delta_{a',p}$. We have $\mbox{Tr}(\Omega_{p^{2}}X_{\rm seed}) = 1\,, \mbox{Tr}(\Omega_{1}X_{\rm seed}) = 0 \,, \mbox{Tr}(\Omega_{p}X_{\rm seed}) = 1$ so in this case the Poincar\'e sum is proportional to:
\begin{equation}
(p^{2}-p)Z_{p^{2}} -p\,Z_{1} + (p^{2}- p+1) Z_{p} = p(p-2)Z_{p^{2}} + (p^{2}- p+1) Z_{p}
\end{equation}
This is true for any seed primary $\abs{\chi_{a}}^{2}$ where $p\mid a$ unless $a=0$.
\end{itemize}
These results can now be compared with the explicit examples in Table \ref{sunone.2} that correspond to $N=p^2$ and we find perfect agreement. 

From this example we learn the general lesson that \poin sums will mix the partition function of a meromorphic CFT (as long as it has a Kac-Moody algebra) with the diagonal invariant of the same Kac-Moody algebra. This potentially leads to many more cases that can be investigated, a point to which he hope to return in the future. 

\subsubsection*{Two invariants: Coefficients for $N=2p^2$}

Here, $m=\frac{N}{2}=p^{2}$, so again the relevant matrices are $\Omega_{p^{2}},\, \Omega_{1},$ and $ \Omega_{p}$. The matrix elements are:
\begin{equation}
(\Omega_{p^{2}})_{aa'} = \delta_{a'a} \,,\quad (\Omega_{1})_{aa'} = \delta_{a',-a} \,,\quad (\Omega_{p})_{aa'} = \begin{cases} 0, \hspace{2.1cm} p \nmid a \mbox{ or } p \nmid a'    \\ \sum_{\xi=0}^{p-1} \delta_{a', a + 2\xi p}, \quad \mbox{otherwise} \end{cases}
\end{equation}
For the case of $\Omega_{p}$, we have used the fact that $\alpha=[p,p]=p$, $N/\alpha^{2}=2$ and so we take $\omega(p)=1$.
The modular invariant functions that we get from these matrices are
\begin{align}
Z_{p^{2}} &= \sum_{a,a'=0}^{N-1}\bar{\chi}_{a} \left( \Omega_{p^{2}} \right)_{aa'} \chi_{a'} = \sum_{a,a'=0}^{N-1}\bar{\chi}_{a}\delta_{a'a}\chi_{a'} = \sum_{a=0}^{N-1}\abs{\chi_{a}}^{2} \\
Z_{1} &= \sum_{a,a'=0}^{N-1}\bar{\chi}_{a} \left( \Omega_{1} \right)_{aa'} \chi_{a'} = \sum_{a,a'=0}^{N-1}\bar{\chi}_{a}\delta_{a',-a}\chi_{a'} = \sum_{a=0}^{N-1}\bar{\chi}_{a}\chi_{-a} \\
Z_{p} &= \sum_{a,a'=0}^{N-1}\bar{\chi}_{a} \left( \Omega_{p} \right)_{aa'} \chi_{a'} = \sum_{\substack{a,a'=0 \\ p|a \, , \, p|a'}}^{N-1}\sum_{\xi=0}^{p-1} \bar{\chi}_{a} \,  \delta_{a', a + 2\xi p} \,  \chi_{a'} = \abs{ \sum_{\substack{l=0}}^{p-1}\chi_{2lp} }^{2} + \abs{ \sum_{\substack{l=0}}^{p-1}\chi_{(2l+1)p} }^{2}
\end{align}
Again $Z_1=Z_{p^2}$, but $Z_p$ is no longer a meromorphic CFT -- rather, it is a two-character CFT.

The elements of the matrix of inner products $d_{JK}$ are:
\begin{align}
&\mbox{Tr}(\Omega_{p^{2}}^{2}) = 2p^{2} \,,\quad \mbox{Tr}(\Omega_{1}^{2}) = 2p^{2} \,,\quad \mbox{Tr}(\Omega_{p}^{2}) = 2p^{2} \\
\mbox{Tr}(\Omega_{p^{2}}\Omega_{1}) = &\mbox{Tr}(\Omega_{1}) = 2 \,,\quad \mbox{Tr}(\Omega_{p^{2}}\Omega_{p}) = \mbox{Tr}(\Omega_{p}) = 2p \,,\quad \mbox{Tr}(\Omega_{1}\Omega_{p}) = 2p
\end{align}
and the linear combinations in the Poincar\'e sums are:
\begin{gather}
\begin{split}
\begin{pmatrix} c_{p^{2}}\\ c_{1} \\ c_{p} \end{pmatrix}
&= \abs{\Gamma_{\rm sub}\backslash \Gamma}
\begin{pmatrix} 2p^{2} & 2 & 2p \\ 2 & 2p^{2} & 2p \\ 2p & 2p & 2p^{2}   \end{pmatrix}^{-1}
\begin{pmatrix} \mbox{Tr}(\Omega_{p^{2}}X_{\rm seed}) \\ \mbox{Tr}(\Omega_{1}X_{\rm seed}) \\ \mbox{Tr}(\Omega_{p}X_{\rm seed})  \end{pmatrix} \\
&= \frac{\abs{\Gamma_{\rm sub}\backslash \Gamma}}{2p^2(p^2-1)}
\begin{pmatrix} p^{2} & 0 & -p \\ 0 & p^{2} & -p \\ -p & -p & 1+p^{2}  \end{pmatrix} \begin{pmatrix} \mbox{Tr}(\Omega_{p^{2}}X_{\rm seed}) \\ \mbox{Tr}(\Omega_{1}X_{\rm seed}) \\ \mbox{Tr}(\Omega_{p}X_{\rm seed})  \end{pmatrix} \\
&= \frac{\abs{\Gamma_{\rm sub}\backslash \Gamma}}{2p^2(p^2-1)}
\begin{pmatrix} p^{2}\,\mbox{Tr}(\Omega_{p^{2}}X_{\rm seed}) - p\,\mbox{Tr}(\Omega_{p}X_{\rm seed})  \\ p^{2}\,\mbox{Tr}(\Omega_{1}X_{\rm seed}) - p\,\mbox{Tr}(\Omega_{p}X_{\rm seed})  \\   - p\,\mbox{Tr}(\Omega_{p^{2}}X_{\rm seed})  - p\,\mbox{Tr}(\Omega_{1}X_{\rm seed}) + (1+p^{2})\mbox{Tr}(\Omega_{p}X_{\rm seed})   \end{pmatrix} \\
\end{split}
\end{gather}
For the various different seeds, this gives:
\begin{itemize}
\item 
$(X_{\rm seed})_{aa'} = \delta_{a0}\delta_{a'0}$. We have $\mbox{Tr}(\Omega_{p^{2}}X_{\rm seed}) = \mbox{Tr}(\Omega_{1}X_{\rm seed}) = \mbox{Tr}(\Omega_{p}X_{\rm seed}) = 1$ and the Poincar\'e sum is proportional to:
\begin{equation}
p(p-1)(Z_{p^{2}} + Z_{1}) + (p-1)^{2}Z_{p} = 2(p-1)\left( p\, Z_{p^{2}} + \frac{(p-1)}{2}Z_{p} \right)
\end{equation}
where we again used $Z_{p^{2}}=Z_{1}$. This will also be true of the seed $\abs{\chi_{p^{2}}}$.
\item 
$(X_{\rm seed})_{aa'} = \delta_{a1}\delta_{a'1}$. We have $\mbox{Tr}(\Omega_{p^{2}}X_{\rm seed}) = 1\,, \mbox{Tr}(\Omega_{1}X_{\rm seed}) = 0 \,, \mbox{Tr}(\Omega_{p}X_{\rm seed}) = 0$ and the Poincar\'e sum is proportional to:
\begin{equation}
p^{2}Z_{p^{2}} - p Z_{p} = p( p\,Z_{p^{2}} - Z_{p})
\end{equation}
This is true for any seed primary $\abs{\chi_{a}}^{2}$ where $p\nmid a$.
\item 
$(X_{\rm seed})_{aa'} = \delta_{ap}\delta_{a'p}$. We have $\mbox{Tr}(\Omega_{p^{2}}X_{\rm seed}) = 1\,, \mbox{Tr}(\Omega_{1}X_{\rm seed}) = 0 \,, \mbox{Tr}(\Omega_{p}X_{\rm seed}) = 1$ and the Poincar\'e sum is proportional to
\begin{equation}
(p^{2}-p)Z_{p^{2}} -p\,Z_{1} + (p^{2}- p+1) Z_{p} = p(p-2)Z_{p^{2}} + (p^{2}- p+1) Z_{p}
\end{equation}
This is true for any seed primary $\abs{\chi_{a}}^{2}$ where $p\mid a$ unless $a=0,p^2$.
\end{itemize}
These predictions agree with the results of explicit computations, of which a few cases are displayed in Table \ref{sunone.2}.

\subsubsection{Poincar\'e Sums for \suthk}

Next we turn to the case of \suthk. These WZW models are considerably more complicated than those for \sutk and \sunone. The classification of modular invariants in this case has been carried out by Gannon \cite{Gannon:1992ty, Gannon:1994cf} but not quite with the same methods as those employed in the classifications for \sutk and \sunone. In particular we do not have the analogue of the matrices $\Omega_\delta$ discussed above in Eqs.(\ref{Omegasutwok}), (\ref{Omegasunone}). This makes it more difficult to make general predictions about the result of the \poin sums, though evaluating these sums for any specific case is straightforward.

The classification of \cite{Gannon:1992ty, Gannon:1994cf} provides a list of physical invariants that appear at different levels. It is found that two invariants called $Z_A, Z_D$ exist for all heights $n=k+3$ (with $k>1$), in contrast to the $SU(2)$ case where $Z_D$ existed only for even $n$. There are also four exceptional invariants denoted $Z_E$ which appear at $n=5,9,21$ (with two distinct invariants appearing at $n=9)$. 

We calculated the Poincar\'e sums for the seed corresponding to each individual character for the values $k=1,2,\dots,16$ using SageMath. As in the $SU(2)_k$ case, we found that the Poincar\'e sums are generally linear combinations of multiple modular invariants. Moreover these combinations often contain unphysical invariants that we have denoted $Z_{new}$. The results up to $k=8$ are given in Table \ref{SU3fulltable}. 
    
    \begin{longtable}{|>{\centering}p{2cm}|>{\centering}p{6cm}|>{\centering}p{6cm}|}

    \hline
    
     \hline
     \textbf{Level} $(n, c, N_{0})$ & \textbf{Seed primary} $(\lambda_1\lambda_2)$ & \textbf{Poincar\'{e} sum} \tabularnewline
     \hline
     \hline
     
     $(4,2,12)$ & $(1 1)$ & $24Z_A$ \tabularnewline
     & $(1 2)$ & $12Z_A$ \tabularnewline
     \hline
     
     $(5,\frac{16}{5},15)$ & $(1 1),(2 2)$ & $24Z_A$ \tabularnewline
     & $(1 2),(1 3)$ & $12Z_A$ \tabularnewline
     \hline
     
     $(6,4,18)$ & $(1 1)$ & $18Z_A$ \tabularnewline
     & $(1 2),(1 3),(2 3)$ & $4.5(3Z_A-Z_D)$ \tabularnewline
     & $(1 4)$ & $4.5(Z_A+Z_D)$ \tabularnewline
     & $(2 2)$ & $18Z_D$ \tabularnewline
     \hline
     
     $(7,\frac{32}{7},21)$ & $(1 1),(2 2),(3 3)$ & $12(Z_A+Z_D)$ \tabularnewline
     & $(1 2),(2 4)$ & $3(9Z_A-7Z_D)$ \tabularnewline
     & $(1 3),(1 4),(1 5),(2 3)$ & $6(Z_A+Z_D)$ \tabularnewline
     \hline
     
     $(8,5,24)$ & $(1 1),(3 3)$ & $4(2Z_A+2Z_D+Z_E)$ \tabularnewline
     & $(1 2),(1 5)$ & $4(4Z_A-2Z_D-Z_E)$ \tabularnewline
     & $(1 3),(3 4)$ & $6(2Z_A-2Z_D+Z_E)$ \tabularnewline
     & $(1 4)$ & $12Z_E$ \tabularnewline
     & $(1 6),(2 3)$ & $2(2Z_A+2Z_D+Z_E)$ \tabularnewline
     & $(2 2)$ & $16(Z_A+Z_D-Z_E)$ \tabularnewline
     & $(2 4),(2 5)$ & $8(Z_A+Z_D-Z_E)$ \tabularnewline
     \hline

     $(9,\frac{16}{3},27)$ & $(1 1),(2 2),(4 4)$ & $3(6Z_A+Z_D)$ \tabularnewline
     & $(1 2),(1 3),(1 5),(1 6),(2 3),$ & $4.5(3Z_A-Z_D)$
     \tabularnewline
     & $(2 4),(2 6),(3 4),(3 5)$ & \tabularnewline
     & $(1 4),(1 7),(2 5)$ & $1.5(3Z_A+5Z_D)$ \tabularnewline
     & $(3 3)$ & $27Z_D$ \tabularnewline
     \hline
     
     $(10,\frac{28}{5},30)$ & $(1 1),(3 3)$ & $4(3Z_A+2Z_D+2Z_{new})$ \tabularnewline
     & $(1 2),(1 3),(1 6),(2 7)$ & $0.5(21Z_A-13Z_D-4Z_{new})$
     \tabularnewline
     & $(1 4),(2 5)$ & $3(3Z_A+Z_D+4Z_{new})$
     \tabularnewline
     & $(1 5),(2 3),(3 5),(4 5)$ & $0.5(27Z_A-15Z_D+6Z_{new})$
     \tabularnewline
     & $(1 7),(3 6)$ & $3Z_A+5Z_D-4Z_{new}$
     \tabularnewline
     & $(1 8),(3 4)$ & $2(3Z_A+2Z_D+2Z_{new})$
     \tabularnewline
     & $(2 2),(4 4)$ & $12(Z_D-2Z_{new})$
     \tabularnewline
     & $(2 4),(2 6)$ & $6(Z_D-2Z_{new})$
     \tabularnewline
     \hline
     
     $(11,\frac{64}{11},33)$ & $(1 1),(2 2),(3 3),(4 4),(5 5)$ & $12(Z_A+Z_D)$ \tabularnewline
     & $(1 2),(1 3),(1 6),(1 8),(2 3),(2 4),$ & $3(5Z_A-3Z_D)$
     \tabularnewline
     & $(2 6),(3 7),(4 5),(4 6)$ & \tabularnewline
     & $(1 4),(1 5),(1 7),(1 9),(2 5),(2 7),$ & $1.5(3Z_A+5Z_D)$ \tabularnewline
     & $(2 8),(3 4),(3 5),(3 6)$ & \tabularnewline
     \hline
     
    \caption{Poincar\'e sums for $SU(3)_k$.}
    \label{SU3fulltable}
    \end{longtable}   
    
Here at height $n=10$, $Z_{new}$ is an unphysical invariant given by:
\be
\begin{split}
Z_{new} &=-|\chi_{12}|^2-|\chi_{13}|^2+|\chi_{14}|^2 -|\chi_{16}|^2-|\chi_{17}|^2-|\chi_{22}|^2\\
&\quad -2|\chi_{24}|^2+|\chi_{25}|^2-2|\chi_{26}|^2-|\chi_{27}|^2-|\chi_{36}|^2-|\chi_{44}|^2\\
&\quad\{-\bchi_{11}\chi_{36}
+\bchi_{12}\chi_{34}
-\bchi_{13}\chi_{18}
+\bchi_{15}\chi_{45}\\
&\quad
-\bchi_{16}\chi_{18}
+\bchi_{17}\chi_{33}
+\bchi_{23}\chi_{35}
+\bchi_{27}\chi_{34}
+{\rm c.c.} \}
\end{split}
\ee    
    
Based on the table we make the following observations:
    \begin{itemize}
        \item Within a given model, several distinct seeds give the same linear combination of modular invariants.
        \item The vacuum seed always gives a linear combination involving non-negative weights.
                \item Unlike the $SU(2)$ case, we have observed that the linear combination for all $n>8$ (up to 16) contains an unphysical invariant.
        \item However whenever the height $n$ is a prime $p>5$, the linear combination from the vacuum seed is proportional to $Z_A+Z_D$ and moreover does not contain any unphysical invariants. 
    \end{itemize}
    
It would be interesting to prove the above statements in general. We will not do so here,  instead we will derive the coefficients for the specific case of $k=4$ and leave a more detailed analysis for future work. From the table we see that at $k=4$, we only find the $Z_{A}$ and $Z_{D}$ invariant. These invariants are given as:
\begin{align}
    Z_{A} &= \sum_{\lambda} \abs{\chi_{\lambda}}^{2}\\
    Z_{D} &= \sum_{\lambda} \bar{\chi}_{A^{kt(\lambda)}\lambda}\,\chi_{\lambda}
\end{align}
where $t(\lambda)=\lambda_{1}-\lambda_{2}$ and $A \, \lambda = (n-\lambda_{1}-\lambda_{2},\lambda_{1})$. The operator $A$ is of order three: $A^{3}=1$. We also have the corresponding conjugate versions of these given by:
\begin{align}
    Z_{A}^{C} &= \sum_{\lambda} \bar{\chi}_{\lambda}\,\chi_{C\lambda} \\
    Z_{D}^{C} &= \sum_{\lambda} \bar{\chi}_{A^{kt(\lambda)}\lambda}\,\chi_{C\lambda}
\end{align}
where $C\lambda$ is the complex conjugate representation to $\lambda$.
This permits us to work out the corresponding $\Omega$ matrices, which turn out to be:
\begin{equation}
(\Omega_{A})_{\lambda,\lambda'} = \delta_{\lambda,\lambda'},~
(\Omega_{CA})_{\lambda,\lambda'} = \delta_{C\lambda,\lambda'},~ 
(\Omega_{D})_{\lambda,\lambda'} = \delta_{\lambda,A^{kt(\lambda')}\lambda'},~ 
(\Omega_{CD})_{\lambda,\lambda'} = \delta_{C\lambda,A^{kt(\lambda')}\lambda'}
\end{equation}
The relevant traces of products of these matrices are then:
\begin{equation}
\begin{split}
\mbox{Tr}(\Omega_{A}^{2}) &= \mbox{Tr}(\Omega_{CA}^{2}) = \mbox{Tr}(\Omega_{D}^{2}) =\mbox{Tr}(\Omega_{CD}^{2}) = 15\\
\mbox{Tr}(\Omega_{A}\Omega_{CA}) &= 3, \quad
\mbox{Tr}(\Omega_{A}\Omega_{D}) = 5, \quad
\mbox{Tr}(\Omega_{A}\Omega_{CD}) = 9, \quad \\
\mbox{Tr}(\Omega_{CA}\Omega_{D}) &= 9, \quad
\mbox{Tr}(\Omega_{CA}\Omega_{CD}) = 5, \quad
\mbox{Tr}(\Omega_{D}\Omega_{CD}) = 3
\end{split}
\end{equation}
Following the usual strategy, we now invert the relevant $4\time4$ matrix to get the linear combinations:
\be
\begin{pmatrix} c_{A}\\ c_{CA} \\ c_{D} \\ c_{CD} \end{pmatrix}
= \frac{\abs{\Gamma_{\rm sub}\backslash \Gamma}}{128}
\begin{pmatrix} 15 a + 3 b - 5 c - 9 d \\ 3 a + 15 b - 9 c - 5 d \\ - 5 a - 9b + 15c + 3d  \\  -9 a - 5 b + 3c + 15d  \end{pmatrix}
\ee
where for compactness we have introduced $a = \mbox{Tr}(\Omega_{A}X_{\rm seed}),~ b = \mbox{Tr}(\Omega_{CA}X_{\rm seed}),~ c = \mbox{Tr}(\Omega_{D}X_{\rm seed}),~ d = \mbox{Tr}(\Omega_{CD}X_{\rm seed})$. 

Let us now consider different choices of seeds:
\begin{itemize}
\item $(X_{\rm seed})_{\lambda\lambda'} = \delta_{\lambda,(1,1)}\delta_{\lambda',(1,1)}$. We have then $a=b=c=d= 1$ and so the Poincar\'e sum is proportional to
\begin{equation}
4(Z_{A} + Z^{C}_{A} + Z_{D} + Z^{C}_{D} ) = 8(Z_{A} + Z_{D})
\end{equation}
where we got the second form after using $Z_{i}=Z_{i}^{C}$.
\item $(X_{\rm seed})_{\lambda\lambda'} = \delta_{\lambda,(1,2)}\delta_{\lambda',(1,2)}$. We have $a=1,b=c=d=0$ and so the Poincar\'e sum is proportional to:
\begin{equation}
15Z_{A} + 3Z^{C}_{A} - 5Z_{D} - 9Z^{C}_{D}  = 2(9Z_{A} - 7Z_{D})
\end{equation}
\item $(X_{\rm seed})_{\lambda\lambda'} = \delta_{\lambda,(1,3)}\delta_{\lambda',(1,3)}$. We have $a=d=1,b=c=0$ and so the Poincar\'e sum is proportional to:
\begin{equation}
6Z_{A} - 2Z^{C}_{A} - 2Z_{D} + 6Z^{C}_{D} = 4(Z_{A} + Z_{D})
\end{equation}
\end{itemize}
These results match precisely with the output of the SageMath calculations presented in the table for this case.

\subsection{Virasoro Minimal Models}

\poin sums for minimal models were first discussed in \cite{Castro:2011zq}. In this case too, all the possible modular invariants are classified \cite{Cappelli:1986hf, Cappelli:1987xt}. Indeed the results are very similar to those for \sutk that we have already encountered.

The characters for minimal models are labelled $\chi_{r,s}$, the range of these being defined in Appendix \ref{basicdata}. The identity character is $\chi_{1,1}$. Modular invariants for the $(m,m')$ minimal model, denoted $Z_{\delta\delta'}$, are defined in terms of a matrix $\Omega_{\delta\delta'}$ by:
\begin{equation}
Z_{\delta\delta'} = \frac{1}{8}\sum_{r_{1},r_{2}=1}^{2m'}\sum_{s_{1},s_{2}=1}^{2m} \bar{\chi}_{r_{1},s_{1}}\, \left[\Omega_{\delta\delta'}\right]_{(r_{1},s_{1}),(r_{2},s_{2})} \, \chi_{r_{2},s_{2}}
\label{doubledZ}
\end{equation}
where $\delta |m$ and $\delta' |m'$. The parameters $m$ and $m'$ act as the analogues of the height $n$ for \sutk, while the matrices $\Omega_{\delta\delta'}$ are simply tensor products of two of the $\Omega_{\delta}$ that we encountered in the $SU(2)_{k}$ case:
\begin{equation}
\left[\Omega_{\delta\delta'}\right]_{(r_{1},s_{1}),(r_{2},s_{2})} = (\Omega_{\delta})_{s_{1},s_{2}}(\Omega_{\delta'})_{r_{1},r_{2}}
\end{equation}
The characters are given as:
\begin{equation}
\chi_{r,s}(\tau) = \frac{1}{\eta(\tau)}\sum_{t=-\infty}^{\infty} \left(q^{\frac{(2tmm' + rm - sm')^{2}}{4mm'}} - q^{\frac{(2tmm' + rm + sm')^{2}}{4mm'}} \right)
\end{equation} 
from which it follows that $\chi_{r,s}=-\chi_{-r,s}=\chi_{r+2m',s}$ and similarly for the shifts of $s$. With this, many of the results for \sutk can be adapted. 

\subsubsection*{One physical invariant}

This is only possible when $\sigma(m)$ and $\sigma(m')$ are each individually equal to $2$ or $3$. This condition is satisfied only if $m$ and $m'$ are either a prime or the square of a prime. This leads to four possibilities: $(m,m') = (p,q), (p^{2},q), (p,q^{2}), (p^{2},q^{2})$, where $p,q$ are distinct primes. Since the diagonal invariant always exists and is physical, it follows that whenever there is a unique modular invariant, it is physical. The corresponding invariant is labelled $Z_{AA}$. However only a couple of unitary minimal models belong in the families listed above, namely $(m,m+1)=(3,4)$ and $(4,5)$. 

\subsubsection*{Two physical invariants}

There will be exactly two modular invariants when one of $m$ or $m'$ is of the form $p$ or $p^{2}$, while the other one is $qr, q^{3}, q^{4}$, where $p,q,r$ are primes. However this does not ensure that both are physical. From our previous discussion we know that physical invariants only occur if $\Omega_{2}$ contributes, which in turn requires $m'\in \{2q, 8,16\}$. Meanwile $m=p,p^2$ as before. Thus the models that contain precisely two physical invariants are those with $(m,m') = (p,2q), (p^{2},2q)$ where $p,q>2$, along with $(m,m') = (p,8), (p^{2},8), (p,16), (p^{2},16)$. The corresponding invariants are labelled $Z_{AA},Z_{AD}$. Due to the symmetry between $m$ and $m'$, we identify models where the values of these integers are interchanged. As already noted in Appendix of \cite{Castro:2011zq}, there are some unitary models in this set -- namely those with $m=2p$ where $p$ is a prime such that $2p+1$ is also prime -- these are known as Sophie Germain primes.

In Table \ref{virmin} one finds that for the unitary minimal models with $(m,m+1)$ ranging from $(5,6)$ to $(10,11)$, only the two physical invariants $Z_{AA}$ and $Z_{AD}$ appear. From the discussion above there are infinitely many more such examples, both unitary and non-unitary. 

Let us see how to derive the linear combination of the two invariants that appears when we perform the Poincar\'e sum for the minimal models of type $(m,m')=(m,2q)$ where $m=p$ or $p^2$. For this we consider the matrices $\Omega_{mm'}$ and $\Omega_{m 2}$ acting on the characters $\chi_{r,s}$ where $1\le r\le m'-1$ and $1\le s\le m-1$. Recall that in the formula \eref{doubledZ} the range of these indices is temporarily doubled, with a constraint that ultimately reduces them to their physical range. For, the case of $\Omega_{mm'}$, we have:
\begin{equation}
\begin{split}
Z_{mm'} &= \frac{1}{8}\sum_{r_{1},r_{2}=1}^{2m'}\sum_{s_{1},s_{2}=1}^{2m} \bar{\chi}_{r_{1},s_{1}}\, \left[\Omega_{mm'}\right]_{(r_{1},s_{1}),(r_{2},s_{2})} \, \chi_{r_{2},s_{2}} \\
&= \frac{1}{8}\sum_{r_{1},r_{2}=1}^{2m'}\sum_{s_{1},s_{2}=1}^{2m} \bar{\chi}_{r_{1},s_{1}}\, \delta_{s_{1},s_{2}}\delta_{r_{1},r_{2}} \, \chi_{r_{2},s_{2}} 
= \frac{1}{2} \sum_{r_{1}=1}^{m'-1}\sum_{s_{1}=1}^{m-1} \abs{\chi_{r_{1},s_{1}}}^{2} 
\end{split}
\end{equation}
Thus when acting on $\chi_{r,s}$, the matrix $(I_{mm'})_{(r_{1},s_{1}),(r_{2},s_{2})} = \delta_{r_{1},r_{2}}\delta_{s_{1},s_{2}}$. Next, we look at the case of $\Omega_{m2}$:
\begin{equation}
\begin{split}
Z_{m2} &= \frac{1}{8}\sum_{r_{1},r_{2}=1}^{2m'}\sum_{s_{1},s_{2}=1}^{2m} \bar{\chi}_{r_{1},s_{1}}\, \left[\Omega_{m2}\right]_{(r_{1},s_{1}),(r_{2},s_{2})} \, \chi_{r_{2},s_{2}} \\
&= \frac{1}{4} \sum_{s_{1}=1}^{m-1} \sum_{r_{1}}^{2m'} \bar{\chi}_{r_{1},s_{1}} \chi_{(m'-1)r_{1},s_{1}} \\
&= \frac{1}{4} \sum_{s_{1}=1}^{m-1} \left( -\sum_{r_{1}\, {\rm even}=1}^{2m'} \abs{\chi_{r_{1},s_{1}}}^{2} + \sum_{r_{1}\, {\rm odd}=1}^{2m'} \bar{\chi}_{r_{1},s_{1}} \chi_{m'-r_{1},s_{1}} \right) \\
\end{split}
\end{equation}
So the matrix $I_{m2}$ is:
\begin{equation}
[I_{m2}]_{(r_{1},s_{1}),(r_{2},s_{2})} = \delta_{s_{1},s_{2}}\times \begin{cases} -\delta_{r_{1},r_{2}}, \hspace{1cm} r_{1} \mbox{ even} \\ ~~\delta_{r_{2},m'-r_{1}},\quad\, r_{1} \mbox{ odd}   \end{cases}
\end{equation}

Now we compute the inner products for these two matrices to get, for $m'=2q$:
\begin{equation}
\begin{split}
\mbox{Tr}(I_{mm'}^{2}) &  = (m-1)(2q-1) \\
\mbox{Tr}(I_{mm'}I_{m2}) &= (m-1)(2-q)\\
\mbox{Tr}(I_{m2}^{2}) &= (m-1)(2q-1) 
\end{split}
\end{equation}
We see that all the inner products are the same as in (\ref{innprod}) upto an overall factor of $(m-1)$ and so we see that the coefficients will be given by solving the same equation as in (\ref{mateqn}). Due to the relation $(r,s)\sim(m'-r,m-r)$, at first it might seem that there is some ambiguity in choosing the $X_{\rm seed}$. However this is not the case. If we want the seed to be $\abs{\chi_{r_{0},s_{0}}}^{2}$, it is easily verified that the two choices: 
\begin{equation}
\begin{split}
\label{4151}
(X_{\rm seed})_{(r,s),(r',s')} &= \delta_{r,r_{0}}\delta_{r',r_{0}}\delta_{s,s_{0}}\delta_{s',s_{0}}\\
(\tilde{X}_{\rm seed})_{(r,s),(r',s')} &= \delta_{r,m'-r_{0}}\delta_{r',m'-r_{0}}\delta_{s,m-s_{0}}\delta_{s',m-s_{0}}
\end{split}
\end{equation}
give the same result. 

The result of this computation is that the linear combinations of physical invariants appearing in the Poincar\'e sums for $(m,2q)$ minimal models (where $m=p$ or $p^{2}$) are exactly the same as those that were obtained in the computations of Poincar\'e sums for the case of $SU(2)_{k}$ where $k+2=n=2q$. This can be verified for the examples in Table \ref{virmin}.

\subsubsection*{Three physical invariants}

Minimal models can have at most three physical invariants. This happens when the conditions are satisfied to admit an exceptional invariant, and requires $m$ to be 12,18 or 30 and $m'$ is any odd number (or the equivalent with $m\leftrightarrow m'$). However almost all such cases have additional unphysical invariants. To have precisely three physical and no unphysical invariants, the total number of invariants should also be 3. This will happen when $m=12,18$ and $m'=p,p^2$ for prime $p$. While there is an exceptional invariant at $m=30$, it is easy to verify that the total number of invariants in that case is at least 4.

Applying this to the unitary models, we find that the cases $(11,12)$ and $(12,13)$ have precisely three physical invariants. At $(13,14)$ we again have only two invariants. Starting with $(14,15)$ we start to encounter three invariants but only for $(17,18),(18,19)$ are there no more. So for $m\ge 14,m\ne 17,18$, and other than the cases discussed in the previous subsection, there is at least one unphysical invariant. This agrees with the results found in \cite{Castro:2011zq} and provides a more complete picture. 

\section{\poin Sums for Multiple Genus-1 Boundaries}

\label{highergenus}

\subsection{Generic RCFT}

In this Section we consider RCFT's on 3-manifolds with multiple genus-1 boundaries. This is crucial for the interpretation that a bulk gravity is dual to an average over CFT's. Suppose we have $N_B>1$ genus-1 boundaries having independent modular parameters $\tau^A, A=1,2,\cdots,N_B$. A generic seed would be made up of linear combinations of terms like:
\be
Z_{i_1j_1}(\tau^1)Z_{i_2j_2}(\tau^2)\cdots Z_{i_{N_B}j_{N_B}}(\tau^{N_B})
\ee
with each factor defined as in Eqs. (\ref{diagseed}) or (\ref{nondiagseed}). Thereafter the \poinsum is carried out over independent modular transforms of each $\tau^A$ \footnote{Alternatively one could identify the modular parameters of all the boundaries and carry out a single \poinsum over them. We do not consider this case here.}. Since we allow linear combinations of seeds as above, the final \poinsum is not a simple product of \poin sums over the contribution from each boundary. Now, once we have determined the measure for averaging over CFT's in the single-boundary case, the same measure should apply in the multi-boundary case. This requires that:
\be
Z(\tau^1,\tau^2,\cdots,\tau^{N_B})=\sum_J c_J\,
Z_J(\tau^1)\,Z_J(\tau^2)\cdots Z_J(\tau^{N_B})
\label{multbound}
\ee
for every $N_B$, where the LHS is the result of the multiple \poinsum described above and the RHS has the {\em same} coefficients $c_J$ that appeared in \eref{avgc}. As emphasised in a different class of examples \cite{Maloney:2020nni}, it is not at all obvious that this can hold for all $N_B>1$. We now turn to an investigation of this criterion.

To start with, consider a 3-manifold with two boundaries $\Sigma$ and $\Sigma'$ of genus $g$ and $g'$, each having its own moduli. If the dual of the bulk gravity theory is a single CFT, then we expect that the one- and two-boundary partition functions should be:
\begin{equation}
Z_{\rm grav}(\Sigma)=Z_{\rm CFT}(\Sigma),\quad
Z_{\rm grav}(\Sigma,\Sigma') = Z_{\rm CFT}(\Sigma)\,Z_{\rm CFT}(\Sigma')
\end{equation}
If the dual of the bulk theory is not a single CFT but rather an ensemble with a probability distribution over physical CFTs, then the gravity partition function will be an average over different CFT partition functions. For one boundary we have seen that:
\begin{equation}
\label{4157}
Z_{\rm grav}(\Sigma) = \langle Z_{\rm CFT}(\Sigma) \rangle = \sum_J c_J\,Z_J(\Sigma), \quad \sum_J c_J = 1 
\end{equation}
where the sum is over physical CFTs. The $J$'th CFT occurs with probability $c_J$ \footnote{Here the $c_J$ are normalised such that their sum is unity, though we have not changed the notation to reflect this. It should be evident that the normalised $c_J$ are equal to the un-normalised ones divided by $\sum_J c_J$.}. Once we have the probability distribution $\{c_J\}$, the two-boundary partition function should be:
\begin{equation}
\label{4158}
Z_{\rm grav}(\Sigma,\Sigma') = \langle Z_{\rm CFT}(\Sigma,\Sigma') \rangle = \sum_{i} c_J\,Z_J(\Sigma)\,Z_J(\Sigma')
\end{equation}
with the same probabilities $c_J$ as in (\ref{4157}). The above equation (\ref{4158}) has an obvious generalisation to an arbitrary number of boundaries, once we know the $c_J$ and $Z_J(\Sigma)$. To show that indeed the dual of bulk gravity theory is an ensemble of CFTs one should be able to compute the LHS and RHS of (\ref{4158}) independently and show that they match. 

Computation of the LHS is complicated for the following reason: one class of multi-boundary manifolds is simply the sum of disconnected single-boundary manifolds. For this choice the \poinsum is simply the product of independent \poin sums over the disconnected components, leading to the result $(\sum_J c_J Z_J(\Sigma))^2$. This is of course not the desired answer. Extra contributions will come from  manifolds that connect two or more boundaries in all possible ways. These can be thought of as wormholes. If we knew the analogue of a ``vacuum'' wormhole (corresponding to some particular seed for the \poinsum) we would know how to compute these contributions and could verify \eref{4158}and its $N$-boundary generalisations. 

Lacking this information, we will take a different approach. We will assume that the contribution of the fully disconnected 3-manifold, each with one genus-1 boundary, is described by a product of independent \poin sums that start with the identity seed, following results in the previous sections. Now we seek a rule for the wormhole contribution corresponding to manifolds with all numbers of boundaries $>1$. We then combine these in all possible ways to generate the full $N_B$-boundary result. Then we try to find a seed such that the \poinsum over it reproduces the desired contribution. We now apply this to the case of \sutk.

\subsection{\sutk WZW Models}

We have seen that the \sutk WZW models with $n=k+2=2p$ (with prime $p>2$) provides an infinite class of models having two physical partition functions. For these models we will be able to find a simple and elegant rule for the ``wormhole'' seeds that precisely reproduces the expected answer. While the details we present below are for \sutk, this procedure works in the same way for all models with two invariants. 

Let us label each boundary torus with its corresponding independent modulus $\tau^{(i)}$. The full partition function should be modular invariant with respect to each of the moduli. To perform the gravity calculation, we compute a Poincar\'e sum analogous to the torus partition function:
\begin{equation}
\label{4159}
Z_{\rm grav}(\tau^{(1)},\tau^{(2)}, \dots ,\tau^{(N)}) = \sum_{\gamma^{(1)}, \dots ,\gamma^{(N)}} Z_{\rm seed}(\gamma^{(1)}\tau^{(1)},\gamma^{(2)}\tau^{(2)}, \dots ,\gamma^{(N)}\tau^{(N)})
\end{equation}

From \eref{idavg} we find that the normalised probabilities, starting with the identity seed, are:
\be
c_A=\frac{p+1}{2p-1},\quad c_{D}=\frac{p-2}{2p-1}
\ee
We would like to show that \eref{4159} is equivalent to:
\begin{equation}
\begin{split}
\label{4161}
Z_{\rm grav}(\tau^{(1)},\tau^{(2)}, \dots ,\tau^{(N)}) &=
\langle Z_{\rm CFT}(\tau^{(1)},\tau^{(2)}, \dots ,\tau^{(N)}) \rangle \\
&= c_A\prod_{i=1}^{N}\,Z_{A}(\tau^{(i)}) + c_D\prod_{i=1}^{N}\,Z_{D}(\tau^{(i)}) \\
\end{split}
\end{equation}
To simplify the notation we will write $Z_{A/D}(\tau^{(i)})$ as $Z_{A/D}^{(i)}$ in what follows. Also for any function that depends on all the $\tau^{(i)}$, we will not write the argument explicitly.

Let us start with the case of two boundaries. Then the postulated answer:
\be
Z_{\rm grav}
= c_A\, Z_{A}^{(1)} Z_{A}^{(2)} 
+ c_D\, Z_{D}^{(1)}Z_{D}^{(2)}
\ee
should come from a sum over manifolds with the two genus-one surfaces as their boundaries. In this case there can be two distinct types of manifold: a pair of disconnected manifolds each with a torus boundary, and a single connected manifold with two disjoint genus-1 boundaries:
\be
Z_{\rm grav}=
Z_{\rm grav}^{\rm disconn.}+Z_{\rm grav}^{\rm conn.}
\ee
From our previous analysis of the single-boundary case, we know that:
\be
Z_{\rm grav}^{\rm disconn.}=\left(c_{A}Z_{A}^{(1)} + c_{D}Z_{D}^{(1)} \right)\left(c_{A}Z_{A}^{(2)} + c_{D}Z_{D}^{(2)} \right)
\ee
It follows that the connected contribution should be:
\begin{equation}
\label{4181}
\begin{split}
Z_{\rm grav}^{\rm conn.} 
&= \left( c_{A}Z_{A}^{(1)}Z_{A}^{(2)} + c_{D}Z_{D}^{(1)}Z_{D}^{(2)} \right) - \left(c_{A}Z_{A}^{(1)} + c_{D}Z_{D}^{(1)} \right)\left(c_{A}Z_{A}^{(2)} + c_{D}Z_{D}^{(2)} \right) \\
&= c_{A}\,c_{D} (Z_{A}^{(1)} - Z_{D}^{(1)}) (Z_{A}^{(2)} - Z_{D}^{(2)}) \\
\end{split}
\end{equation}
Thus the connected part is a product of terms of the form $Z_{A}-Z_{D}$ with an overall factor depending on the probabilities. We will see that this is the general form for the connected contribution linking any number of boundaries. 

Before going to the general case, let us extend the above calculation to three genus-one boundaries. Here we expect that: 
\begin{equation}
\begin{split}
Z_{\rm grav} &= c_{A}\prod_{i=1}^{3}Z_{A}^{(i)} + c_{D}\prod_{i=1}^{3}Z_{D}^{(i)}  \\
\end{split}
\end{equation}
Now we have three distinct types of manifolds to sum over: (i) a sum of three disconnected pieces each with one boundary, (ii) a manifold that connects any two of the three boundaries, together with one that has a single boundary, (iii) manifolds which connect all three boundaries. These cases are associated, respectively, to the following partitions of $N=3$: $\{1,1,1\}, \{2,1\},\{3\}$. Henceforth we will label different contributions by partitions of $N$.
Thus we can write:
\begin{equation}
Z_{\rm grav} = Z_{\rm grav}^{\{111\}} + Z_{\rm grav}^{\{21\}} + Z_{\rm grav}^{\{3\}}
\end{equation}
Note that some of the partitions occur in multiple ways, for example there are three ways to realise $\{2,1\}$ depending on the choice of boundary that remains disconnected.
The first term on the RHS will be a product over three copies of (\ref{idavg}), while the second term will be a sum of products of one copy of (\ref{4181}) and one copy of (\ref{idavg}), in three different ways. Then the last term should be:
\begin{equation}
\begin{split}
Z_{\rm grav}^{\{3\}} &= Z_{\rm grav} - Z_{\rm grav}^{\{111\}} - Z_{\rm grav}^{\{21\}} \\
&= c_A\prod_{i=1}^{3}Z_{A}^{(i)} + c_D \prod_{i=1}^{3}Z_{D}^{(i)}  - \prod_{i=1}^{3}(c_AZ_{A}^{(i)} + c_D Z_{D}^{(i)}) \\
&- c_A\,c_D \, (Z_{A}^{(1)} - Z_{D}^{(1)}) (Z_{A}^{(2)} - Z_{D}^{(2)})(c_AZ_{A}^{(3)} + c_D Z_{D}^{(3)}) \\
&- c_A\,c_D \, (Z_{A}^{(1)} - Z_{D}^{(1)}) (Z_{A}^{(3)} - Z_{D}^{(3)})(c_AZ_{A}^{(2)} + c_D Z_{D}^{(2)}) \\
&- c_A\,c_D \, (Z_{A}^{(2)} - Z_{D}^{(2)}) (Z_{A}^{(3)} - Z_{D}^{(3)})(c_AZ_{A}^{(1)} + c_D Z_{D}^{(1)}) \\
\end{split}
\end{equation}
where the contribution of $Z_{\rm grav}^{\{21\}}$ contains three terms, one for each pair of boundaries that are connected. Upon simplifying we have:
\begin{equation}
Z_{\rm grav}^{\{3\}} = c_Ac_D (c_D -c_A)(Z_{A}^{(1)} - Z_{D}^{(1)})(Z_{A}^{(2)} - Z_{D}^{(2)})(Z_{A}^{(3)} - Z_{D}^{(3)})
\end{equation}
so, we see again that the completely connected part is given as product of terms of the form $Z_{A}-Z_{D}$. 
This is now easily generalised to $N_B$ boundaries. We find:
\begin{equation}
\label{4187}
Z_{\rm grav}^{\{N_B\}} = f(c_A,N_B)\prod_{i=1}^{N_B}(Z_{A}^{(i)} - Z_{D}^{(i)}), \quad N_B>1
\end{equation}
where $f(c_A,N_B)$ is a function of the number of boundaries and the probability $c_A$, which we will compute below (we have used $c_D=1-c_A)$. 

These observations tells us that if we can find a seed for generating $Z_{\rm grav}^{\{N_B\}}$ then we can find the seed for the full $N_B$-boundary gravity partition function. For example,
\begin{equation}
\label{4188}
\begin{split}
Z_{\rm seed}^{N_B=1} &=  Z_{\rm seed}^{\{1\}}(\tau^{(1)}) \\
Z_{\rm seed}^{N_B=2} &=  \prod_{i=1}^2 Z_{\rm seed}^{\{1\}}(\tau^{(i)})+ Z_{\rm seed}^{\{2\}}(\tau^{1},\tau^{(2)}) \\
Z_{\rm seed}^{N_B=3} &= \prod_{i=1}^{3}Z_{\rm seed}^{\{1\}}(\tau^{(i)})
 + Z_{\rm seed}^{\{1\}}(\tau^{(1)})\,Z_{\rm seed}^{\{2\}}(\tau^{(2)},\tau^{(3)}) + 
 Z_{\rm seed}^{\{1\}}(\tau^{(2)})\,Z_{\rm seed}^{\{2\}}(\tau^{(1)},\tau^{(3)}) 
 \\
&\quad + Z_{\rm seed}^{\{1\}}(\tau^{(3)})\,Z_{\rm seed}^{\{2\}}(\tau^{(1)},\tau^{(2)}) 
+ Z_{\rm seed}^{\{3\}}(\tau^{(1)},\tau^{(2)},\tau^{(3)}) \\
\end{split}
\end{equation}
One possible choice for $Z_{\rm seed}^{\{1\}}$, as we have seen, is:
\be
Z_{\rm seed}^{\{1\}}\sim \abs{\chi_{1}}^{2}
\label{Zseedchoice}
\ee
with a suitable normalisation. But as we have seen in  Eqs.\,(\ref{4189}),(\ref{4190}), there are more general choices for this seed. Inserting the normalised values of $c_A,c_D$ for the identity seed, namely $\frac{p+1}{2p-1},\frac{p-2}{2p-1}$, we find the general answer  to be given by:
\begin{equation}
Z^{\{1\}}_{\rm seed} =a_{1} \!\!\!\sum_{\substack{\lambda\, \rm odd,\,  \ne p}}\!\!\! b_{\lambda}\abs{\chi_{\lambda}}^{2}  + a_{2}\!\! \sum_{\substack{\lambda\, \rm even}}\!\!c_{\lambda}\abs{\chi_{\lambda}}^{2}  + a_{3}\abs{\chi_{p}}^{2}
 + a_{4} \sum_{\lambda\, {\rm odd},=1}^{p-2}d_{\lambda}(\bchi_{\lambda}\chi_{2p-\lambda} + {\rm c.c.})
\label{Zseed1} 
\end{equation}
with:
\be
a_3=\frac{1}{2p-1}-a_1-a_2,\quad 2a_4=a_1+2a_2-\frac{1}{2p-1},\quad \sum b_\lambda=\sum c_\lambda = \sum d_\lambda=1
\label{Zseed1.cond}
\ee
Clearly $a_1=\frac{1}{2p-1}, b_1=1$ with all other $a_i,b_\lambda=0$ is a possible solution in this case, leading to \eref{Zseedchoice}. But far more general solutions are allowed. 

Let us now try to find a seed that will give us the linear combination $Z_A - Z_D$. This is given by Eqs.(\ref{4189}),(\ref{4190}) with $\alpha=1,\beta=-1$. Then \eref{4190} leads to:
\be
a_3=\frac{p-2}{3(p^2-1)}-a_1-a_2,\quad
2a_4=a_1+2a_2-\frac{1}{p+1}
\ee
This time there is no solution with all $a_i=0$ other than $a_1$. One of the simplest solutions is:
\be
a_1=\frac{1}{p+1},\quad a_3=\frac{1-2p}{3(p^2-1)},\quad a_2=a_4=0,\quad b_1=1
\ee
However as we will shortly see, there are other solutions that are of interest.

We can now write the connected contribution for $N_B$ boundaries:
\begin{equation}
\label{4192}
Z_{\rm grav}^{\rm conn.}(\tau^{(1)},\dots,\tau^{(N_B)}) = \sum_{\gamma^{(i)}\in\Gamma(N_{0})\backslash\Gamma} Z_{\rm seed}^{\rm conn.}(\{\gamma^{(i)}\tau^{(i)}\})
\ee
where:
\be
Z_{\rm seed}^{\rm conn.}(\{\tau^{(i)}\}) = f(c_A,N_B) \prod_{i=1}^{N_B}Z_{\rm seed}^{A-D}(\tau^{(i)}) 
\label{fdef}
\ee
Finally, using (\ref{4192}) in expressions of the  form (\ref{4188}) gives the general seed for the partition function with $N_B$ genus-one boundaries.

It remains to find the form of the function $f(c_A,N_B)$. 
We provide a recursive algorithm to generate this function, where the answer for a given $N_B$ is given in terms of the answer for all numbers of boundaries $<N_B$. To do this we first introduce a generating function for computing the moments of a random variable (say $x$):
\begin{equation}
P(k) = \langle e^{kx} \rangle = \sum_{n\ge0}\frac{k^{n}}{n!} \langle x^{n} \rangle \quad \mbox{and} \quad \mbox{ln }P(k) \coloneqq \sum_{n\ge1}\frac{k^{n}}{n!}\langle x^{n} \rangle_{c}
\end{equation}
where $\langle x^{n} \rangle$ are the usual moments and $\langle x^{n} \rangle_{c}$ denote the connected moments. We can use the above equations to find relations between the two:
\begin{equation}
\sum_{n\ge0}\frac{k^{n}}{n!} \langle x^{n} \rangle = \exp(\sum_{n\ge1}\frac{k^{n}}{n!}\langle x^{n} \rangle_{c})
= \prod_{n\ge1}\sum_{l\ge0} \frac{k^{nl}}{l!}\left(\frac{\langle x^{n} \rangle_{c}}{n!}\right)^{l}
\ee
which implies that:
\be
\langle x^{m} \rangle = m!\sum_{\{l_{n}\}} \prod_{n\ge1}^{m} \frac{1}{l_{n}!}\left(\frac{\langle x^{n} \rangle_{c}}{n!}\right)^{l_{n}}
\end{equation}
The sum is over the set of $\{l_{n}\}$ such that $\sum_{n}nl_{n}=m$, which are just the partitions of the integer $m$. 

To apply this to the present case we replace $x$ by $Z$ and find:
\begin{equation}
\langle Z^{m} \rangle = m!\sum_{\{l_{n}\}} \prod_{n\ge1}^{m} \frac{1}{l_{n}!}\left(\frac{\langle Z^{n} \rangle_{c}}{n!}\right)^{l_{n}}
\ee
Replacing the LHS by the desired average over products of CFT's, and separating two terms in the RHS from the rest in an obvious way, we get:
\be
(c_AZ_{A}^{m} + c_D Z_{D}^{m})= \langle Z \rangle^{m} + m!\sum_{\{l_{n}\}}' \prod_{n\ge1}^{m-1} \frac{1}{l_{n}!}\left(\frac{\langle Z^{n} \rangle_{c}}{n!}\right)^{l_{n}} + \langle Z^{m} \rangle_{c}
\ee
This can now be solved to get the connected part:
\be
\langle Z^{m} \rangle_{c} = (c_AZ_{A}^{m} + c_D Z_{D}^{m}) - (c_AZ_{A} + c_D Z_{D})^{m} - m!\sum_{\{l_{n}\}}' \prod_{n\ge1}^{m-1} \frac{1}{l_{n}!}\left(\frac{\langle Z^{n} \rangle_{c}}{n!}\right)^{l_{n}}
\end{equation}
Now using the fact that $\langle Z^{m} \rangle_{c}\,, (n>1)$ is of the form $f(c_A,m)(Z_{A}-Z_{D})^{m}$ and comparing the coefficients of $Z_{A}^{m}$, we finally get:
\begin{equation}
f(c_A,m) = c_A(1-c_A^{m-1}) - m!\sum_{\{l_{n}\}}' \frac{(c_A)^{l_{1}}}{l_{1}!} \prod_{n\ge2}^{m-1} \frac{1}{l_{n}!}\left(\frac{ f(c_A,n) }{n!}\right)^{l_{n}}
\end{equation}
This is the desired recursive formula for the coefficients in \eref{fdef}.
Some examples of the above formula are as follows:
\be
\begin{split}
m=2\!:\quad f(c_A,2) &= c_A(1-c_A) \\
m=3\!:\quad f(c_A,3) &= c_A(1-c_A^{2}) - 3(c_A)^{2}(1-c_A) = c_A(1-c_A)(1-2c_A) \\
m=4\!:\quad f(c_A,4) &= c_A(1-c_A)(1-6c_A+6c_A^{2})
\end{split}
\ee

The fact that such a rule exists at all, and has nice properties, is very encouraging. In principle it could be checked if we can have a better understanding of the wormhole-type manifolds, though -- as already emphasised here and in \cite{Maloney:2020nni}, the semi-classical notion of manifold does not really apply in the RCFT context. 

We have shown how to write down a general seed for the partition function with $N_B$ genus-1 boundaries. This expressions involved free parameters and hence give an infinite family of possible seeds. To conclude this section we now give an example of a particular choice of these parameters which allows us to write down the full $N_B$-boundary seed in a simple compact form. We are not sure whether there is any physical interpretation for this choice, but mathematically it is appealing.

We begin with the general form of $Z_{\rm seed}^{\{1\}}$, given in \eref{Zseed1}. Now we set $b_{1}=c_{2}=1$ and all the remaining $b_{\lambda}=c_{\lambda}=0$. We also demand that $a_{4}=0$.  From \eref{Zseed1.cond}, this leads to the constraint $a_{1}+2a_{2}=\frac{1}{2p-1}$, so the disconnected seed can be written:
\begin{equation}
\label{6153}
Z^{\{1\}}_{\rm seed} = \left( \frac{1}{2p-1} - 2a_{2} \right) \abs{\chi_{1}}^{2} + a_{2} \abs{\chi_{2}}^{2} + a_{2} \abs{\chi_{p}}^{2}
\end{equation}
For the case of $Z_{\rm seed}^{A-D}$, we make a similar demand. Here we will denote the parameters with a prime to distinguish from the above. We set $b_{1}', c_{2}'=1$
with the remaining $b_{\lambda}'=c_{i}'=0$. We also set $a'_{4}=0$. This implies $a'_{1} + 2a'_{2} = \frac{1}{p+1}$. Then the corresponding seed is:
\begin{equation}
\label{6154}
Z^{A-D}_{\rm seed} = \left( \frac{1}{p+1} - 2a'_{2} \right) \abs{\chi_{1}}^{2} + a'_{2} \abs{\chi_{2}}^{2} + \left( a'_{2} - \frac{2p-1}{3(p^{2}-1)} \right) \abs{\chi_{p}}^{2}
\end{equation}
Thus we have found the required seeds to set up the recursion relation, as functions of two arbitrary parameters $a_2,a_2'$. We now show that for special values of these parameters, things simplify considerably. Let us choose:
\begin{equation}
\label{6155}
a_{2} = \frac{p-2}{3 (p-1) (2 p-1)} \,,\quad a_{2}' = \frac{p-2}{3 (p-1) (p+1)}
\end{equation}
Now using (\ref{4188}), (\ref{6153}) and (\ref{6155}) we find for one boundary:
\begin{equation}
Z_{\rm seed}^{N_B=1} = Z_{\rm seed}^{\{1\}} = \frac{p+1}{(2 p-1) (3p-3) } \left( \abs{\chi_{1}}^{2} + \frac{p-2}{p+1} \abs{\chi_{2}}^{2} + \frac{p-2}{p+1} \abs{\chi_{p}}^{2} \right)
\end{equation} 
while for two boundaries, we get:
\begin{equation}
\begin{split}
Z_{\rm seed}^{N_B=2} &= \prod_{i=1}^{2} Z_{\rm seed}^{\{1\}}(\tau^{(i)}) + p_{A}(1-p_{A})Z_{\rm seed}^{A-D}(\tau^{(1)},\tau^{(2)}) \\
                             &= \frac{p+1}{(2p-1)(3p-3)^{2}}\left( \prod_{i=1}^{2} \left( \abs{\chi_{1}^{(i)}}^{2} + \left(\frac{p-2}{p+1}\right)\abs{\chi_{2}^{(i)}}^{2} \right)  + \left(\frac{p-2}{p+1}\right) \prod_{i=1}^{2}\abs{\chi_{p}^{(i)}}^{2} \right) \\
\end{split}
\end{equation}
From the above, we see a pattern emerging. So one may conjecture that for $N_B$ boundaries, $Z_{\rm seed}$ is:
\begin{equation}
\label{6159}
Z_{\rm seed}^{N_B} =  \frac{p+1}{(2p-1)(3p-3)^{N_B}}\left( \prod_{i=1}^{N} \left( \abs{\chi_{1}^{(i)}}^{2} + \left(\frac{p-2}{p+1}\right)\abs{\chi_{2}^{(i)}}^{2} \right)  + \left(\frac{p-2}{p+1}\right) \prod_{i=1}^{N_B}\abs{\chi_{p}^{(i)}}^{2} \right)
\end{equation} 
Now one proves the conjecture by directly evaluating the Poincar\'e sum over this seed and finding that it gives the desired $N_B$-boundary answer. We get:
\begin{equation}
\begin{split}
\sum_{\gamma}Z_{\rm seed}^{N_B} &=  \frac{p+1}{(2p-1)(3p-3)^{N_B}}\left( \prod_{i=1}^{N_B} (3p-3)Z_{A}^{(i)} + \left(\frac{p-2}{p+1}\right) \prod_{i=1}^{N_B}(3p-3)Z_{D}^{(i)}  \right) \\
&= \frac{1}{2p-1} \left( (p+1)  \prod_{i=1}^{N_B} Z_{A}^{(i)} + (p-2) \prod_{i=1}^{N_B}  Z_{D}^{(i)} \right) \\
&= \langle Z(\tau^{(1)}) \cdots Z(\tau^{(N_B)}) \rangle \\
\end{split}
\end{equation}
Thus we see that (\ref{6159}) provides a simple, compact choice of seed which reproduces exactly the desired result for the $N_B$-boundary partition function. However, as explained above, it is not unique and one can have more complicated expressions by choosing the parameters differently. It would be worth investigating whether some additional physical criterion determines the free parameters to produce this compact result. 

\section{Discussion and Conclusions}

\label{conclusions}

We have attempted to resurrect the original attempt by \cite{Castro:2011zq} of finding ``non-semi-classical'' 3d gravity duals to 2d rational conformal field theories. While doing so, we have generalised their considerations to \sunk WZW models, and also extended their results on Virasoro minimal models. Most important, we have explored the idea that the dual of AdS$_3$ gravity can be an average of an ensemble of CFT's rather than a single CFT. The ensembles we have studied are extremely small, in most cases having just two members, which makes the problem highly tractable. In the present work, our considerations have been limited to 3d spacetimes with one or more genus-1 boundaries.

Our principal results are: (i) there are infinitely many unitary WZW models having single CFT duals, whose partition functions are correctly reproduced by a \poinsum. This is in contrast to the minimal models, for which the same considerations only lead to two unitary cases with single duals, namely $(m,m+1)=(3,4)$ and $(4,5)$, (ii) there are infinitely many unitary WZW models, as well as both unitary and non-unitary minimal models, whose \poinsum gives an average of two RCFT's. In these cases the coefficient of the average can be computed for the entire infinite series as simple functions of certain prime numbers labelling the individual models in the series, (iii) in the cases where there are two RCFT's, it is possible to define a \poin series that correctly averages the products of partition functions of each RCFT over multiple boundaries. To do so, it requires contributions from (non-semi-classical) manifolds that are both disconnected and that connect multiple boundaries in all possible ways, the latter being some kind of analogue wormholes, (iv) there are also cases, notably for \sunone, where the dual can be an average over arbitrarily many CFT's.

Our analysis has been most detailed for \sutk models. However the results for minimal models are quite similar. This seems to arise from a kind of ``doubling'' relation between the two classes of models, which is well-known from the classic papers of Cappelli, Itzykson and Zuber \cite{Cappelli:1986hf, Cappelli:1987xt}. We hope to return to more general classes of RCFT for which a given set of characters admits more than one modular invariant. 

In the early studies of \poinser for gravity \cite{Maloney:2007ud}, it was assumed that the ``seed'' for the series is the identity module, corresponding to thermal AdS$_3$ on the gravity side. Later studies, notably \cite{Keller:2014xba}, allowed for more general primary seeds and studied their contribution. The recent work \cite{Benjamin:2020mfz} also invokes additional seeds associated to conical singularities. Thus there is no real consensus on what should be the seed for a \poinser. In the present work we have similarly taken an agnostic view on this. However, as seen in many of our tables, the identity seed often seems to return a positive linear combination of CFT partition functions, while other primary seeds often do not \footnote{Amusingly this seems to be the opposite of what is found in pure AdS$_3$ gravity, where the identity seed leads to negative coefficients and other primary seeds are added to try and cure this.}. Related to this is the question of additional seeds that must be added to account for analogue wormholes in the multi-boundary case. We have found candidates that work correctly, which we think is remarkable, but we do not have a first-principles reason why the wormhole seeds should be what we propose. Progress on this issue would be most helpful and could illuminate the more general case of pure AdS gravity.

Our considerations have been based on considering Euclidean AdS$_3$ with one or more genus-1 boundaries. For minimal models, it has been argued \cite{Jian:2019ubz} that stronger constraints are obtained by considering, for example, a single genus-2 boundary. It would be worthwhile to know what constraints can be put on the dualities described in the present work by considering genus-2 or higher-genus boundaries. 

In the context of Narain lattices which have U(1)$^{2D}$ Kac-Moody algebras, \cite{Maloney:2020nni} proposed that the gravity dual is a topological U(1)$^{2D}$ Chern-Simons theory. In the same spirit, we have suggested that the gravity dual for WZW theories based on non-abelian \sunk $\times$ \sunk Kac-Moody algebras is a topological \sunk $\times$ \sunk Chern-Simons theory. However we did not really make use of this in the present work. Perhaps one can obtain more information about the dualities discussed by invoking properties of the ``gravity'' side.

Among possible future directions, one would be to investigate \sunone WZW models in greater detail. Since all the modular invariants are physical, we do not need to restrict to any special class as was the case for \sutk. Also the number of invariants is unbounded. Moreover, for any given number of invariants there will, in general, be an infinite family of models within this category. Another interesting point is that for each such family we can consider the large-$N$ (large-$\mathbf{c}$) limit, which may permit considerations that are closer to being semi-classical. We hope to return to this in future. 

Another fascinating direction is to start with meromorphic RCFT. As is well-known, these correspond to free bosons on an even unimodular lattice, orbifolds thereof and some generalisations \cite{Goddard:1983at, Schellekens:1992db}. Such CFT's typically have Kac-Moody algebras and their single character is just an integral linear combination of Kac-Moody characters. This automatically gives us (at least) a pair of CFT's: the meromorphic one and the diagonal invariant. Thus when computing \poin sums, one expects to find linear combinations of both partition functions. This has already been verified in one example, namely SU(25)$_1$, in the present paper -- see the discussion following \eref{sunoneinv}. A large fraction of meromorphic CFT's are based on non-simple Kac-Moody algebras which renders the analysis more novel as well as more complicated.

Recently, the considerations of \cite{Castro:2011zq} have been extended to the case of boundary minimal models \cite{Karch:2020flx}. This requires the introduction of Randall-Sundrum branes \cite{Randall:1999vf} associated to Cardy states \cite{Cardy:1986gw}. It should be possible to generalise these ideas to the WZW models that we have studied here, and even take the large-$\bc$ limit as suggested above. 

Finally, the MLDE approach to classifying rational CFT \cite{Mathur:1988na, Mukhi:2019xjy} might be fruitfully combined with the \poinsum method to generate new examples of RCFT with multiple modular invariants. 

\section*{Acknowledgements}

We thank Matthias Gaberdiel, Alex Maloney and Rahul Poddar for helpful discussions, and Tushar Gopalka for collaboration at the early stages of this project. VM would like to acknowledge the INSPIRE Scholarship for Higher Education, Government of India. PS would like to acknowledge support from the Clarendon Fund and the Mathematical Institute, University of Oxford. All of us are grateful for support from a grant by Precision Wires India Ltd.\ for String Theory and Quantum Gravity research at IISER Pune.

\appendices

\section*{Appendices}

\section{Congruence subgroups}

\label{congsub}

Here we collect a number of definitions and properties of congruence subgroups of $\Gamma=$ SL(2,$\mathbb{Z}$). We have:
\be
\begin{split}
\Gamma(N)& :=\left\{\begin{pmatrix} a& b\\ c& d\end{pmatrix}\in \Gamma\bigg|~ a,d\equiv 1; b,c\equiv 0 \hbox{ mod } N\right\}\\
\Gamma_1(N)& :=\left\{\begin{pmatrix} a& b\\ c& d\end{pmatrix}\in \Gamma\bigg|~ a,d\equiv 1; c\equiv 0 \hbox{ mod } N\right\}\\
\Gamma_0(N)& :=\left\{\begin{pmatrix} a& b\\ c& d\end{pmatrix}\in \Gamma\bigg|~  c\equiv 0 \hbox{ mod } N\right\}
\end{split}
\ee
We have the inclusions: $\Gamma(N)\subset \Gamma_1(N)\subset \Gamma_0(N)\subset \Gamma$. Also, $\Gamma(nN)\subset \Gamma(N)$ for any integer $n>1$, and similarly for the others.
We also have:
\be
\Gamma^0(N) :=\left\{\begin{pmatrix} a& b\\ c& d\end{pmatrix}\bigg|~  b\equiv 0 \hbox{ mod } N\right\}
\ee

$\Gamma(N)$ is called the principal congruence subgroup, and is a normal subgroup of $\Gamma$. Any subgroup of $\Gamma$ that contains $\Gamma(N)$ for some $N$ is called a congruence subgroup, but will in general not be normal. The smallest such $N$ (corresponing to the largest $\Gamma(N)$) is called the {\em level} of the congruence subgroup. 

We can write down the indices of various congruence subgroups in $\Gamma$:
\be
\begin{split}
[\Gamma:\Gamma(N)]&=N^3\prod_{p|N}\left(1-\frac{1}{p^2}\right)\\
[\Gamma:\Gamma_1(N)]&=N^2\prod_{p|N}\left(1-\frac{1}{p^2}\right)\\
[\Gamma:\Gamma_0(N)]&=N\prod_{p|N}\left(1+\frac{1}{p}\right)
\end{split}
\ee
from which we find the useful relation:
\be
[\Gamma_1(N):\Gamma(N)]=N
\ee

\section{Basic formulae for WZW and Virasoro minimal models}
\label{basicdata}

\subsection*{\sutk WZW models}

We will restrict our attention to the unitary series of \sutk models. An integer level $k$ characterises the values of $\bc,h_i$ in the model. Defining the ``height'' $n=k+2$, we have:
\begin{equation}
\bc = 3-\frac{6}{n} \,, \quad (n {~\rm integer}\ge 3)
\end{equation}
The primaries can be labelled by $\lambda = 2j+1$ where $j$ is the isospin, with $1\le \lambda\le n-1$. Their holomorphic conformal dimensions are given as:
\begin{equation}
h_{\lambda}= \frac{\lambda^{2} - 1}{4n}
\end{equation}
Under modular transformations $T$ and $S$, the characters $\chi_\lambda(\tau)$ go into linear combinations of themselves as follows:
\begin{equation}
\begin{split}
T_{\lambda, \lambda'} &= \delta_{\lambda,\lambda'} \exp(2\pi i \left(h_{\lambda} - \frac{\bc}{24} \right)) \\
S_{\lambda, \lambda'} &= \sqrt{\frac{2}{n}}\sin(\frac{\pi \lambda\lambda'}{n}) \\
\end{split}
\end{equation} 

\subsection*{SU(3)$_k$ WZW models}

For \suthk the representations are labelled by a pair of integers $\lambda_1,\lambda_2\ge 1$. In terms of these and the height $n=k+3$, we have the constraint $\lambda_1+\lambda_2<n$. The central charge and conformal dimensions are:
    \begin{equation}
        \mathbf{c} = 8-\frac{24}{n} \;,\quad h_\lambda = \frac{\lambda_1^2+\lambda_2^2+\lambda_1\lambda_2 - 3}{3n}
    \end{equation}
    It follows from the expression for the conformal dimensions that the semi-conductor in this case has the simple form, $N_0=3n$, whereas the total number of characters that appear at level $k$ is $\frac{n(n-1)}{2}$. Finally, the modular $T$ and $S$ matrices can be expressed as:
\begin{equation}
         	\begin{split}
            T_{\lambda,\mu} &= \delta_{\lambda,\mu} \exp(2\pi i\left(h_\lambda-\frac{c}{24}\right))  \\
            \\
     		S_{\lambda,\mu} &= -\frac{i}{n\sqrt 3} \left[ \exp(\frac{2\pi i}{3n}\left(2\lambda_1\mu_1 + \lambda_1\mu_2 + \lambda_2\mu_1 + 2\lambda_2\mu_2\right)) \right. \\
            &\hspace*{1.6cm} + \exp(\frac{2\pi i}{3n}\left(-\lambda_1\mu_1 - 2\lambda_1\mu_2 + \lambda_2\mu_1 - \lambda_2\mu_2\right)) \\
            &\hspace*{1.6cm} + \exp(\frac{2\pi i}{3n}\left(-\lambda_1\mu_1 + \lambda_1\mu_2 - 2\lambda_2\mu_1 - \lambda_2\mu_2\right)) \\
            &\hspace*{1.6cm} - \exp(\frac{2\pi i}{3n}\left(-\lambda_1\mu_1 - 2\lambda_1\mu_2 - 2\lambda_2\mu_1 - \lambda_2\mu_2\right)) \\
            &\hspace*{1.6cm} - \exp(\frac{2\pi i}{3n}\left(2\lambda_1\mu_1 + \lambda_1\mu_2 + \lambda_2\mu_1 - \lambda_2\mu_2\right)) \\
            &\hspace*{1.6cm} \left. - \exp(\frac{2\pi i}{3n}\left(-\lambda_1\mu_1 + \lambda_1\mu_2 + \lambda_2\mu_1 + 2\lambda_2\mu_2\right)) \right]
        \end{split}
    \end{equation}

\subsection*{SU($N$)$_k$ WZW models}

As in the case of SU(2)$_k$, the level $k$ characterises the values of $\mathbf{c},h_i$ in the model. Defining the height $n=k+N$, we have:
    \begin{equation}
        \mathbf{c} = \frac{k(N^2-1)}{N+k} = (N^{2}-1)\left(1 - \frac{N}{n}\right)
    \end{equation}
    For any given level $k$ only a finite number of representations are allowed, which are selected by the constraint:
    \begin{equation}
    \label{673}
        \sum_{i=1}^{N-1} \lambda_i < n
    \end{equation}
    where $\lambda\equiv(\lambda_1,\lambda_2,\dots,\lambda_{N-1})$ is the Dynkin label for SU($N$) \footnote{Note that we are following the convention where the Dynkin label of the trivial representation is the unit vector $(1,1,\dots,1)$, instead of the zero vector.}, which labels the characters. The number of such representations is given by:
    \begin{equation}
        \#(\text{at level } k) = \frac{(N+k-1)!}{k!(N-1)!}
    \end{equation}
    The conformal dimension of the representation labelled by $\lambda$ can be expressed as:
    \begin{equation}
        h_\lambda = \frac{(\lambda-\rho,\lambda+\rho)}{2n}
    \end{equation}
    where $\rho=(1,1,\dots,1)$, and the inner product is $(x,y) = x_i \kappa_{ij} y_j$, where $\kappa$ is the quadratic form matrix of SU($N$):
    \begin{equation}
        \kappa \equiv \frac{1}{N}
        \begin{pmatrix}
            N-1 & N-2 & N-3 & \cdots & 2 & 1 \\
            N-2 & 2(N-2) & 2(N-3) & \cdots & 4 & 2 \\
            N-3 & 2(N-3) & 3(N-3) & \cdots & 6 & 3 \\
            \vdots & \vdots & \vdots & \ddots & \vdots & \vdots \\
            2 & 4 & 6 & \cdots & 2(N-2) & N-2 \\
            1 & 2 & 3 & \cdots & N-2 & N-1 \\
        \end{pmatrix}
    \end{equation}
The action of $S$ and $T$ modular transformations on the characters are given as:
	\begin{align}
		T_{\lambda\lambda'} &= \delta_{\lambda,\lambda'}\exp(2\pi i \left( h_{\lambda} - \frac{\mathbf{c}}{24}\right)) \\
		S_{\lambda\lambda'} &= \frac{i^{N(N-1)/2}}{\sqrt{Nn^{N-1}}} \sum_{w\in W} \mbox{det}(w)\, \exp(-2\pi i \left( \frac{\lambda \cdot w(\lambda')}{n} \right) )
	\end{align}
	where $W$ is the Weyl group of $SU(N)$.

\subsection*{Virasoro minimal models}

These models are labeled by two co-prime integers $(m,m')$ with $m'>m>2$. The unitary minimal models are the sub-series obtained by setting $m'=m+1$. The central charge in terms of these parameters is:
\begin{equation}
c(m,m') = 1 - \frac{6(m-m')^{2}}{mm'}
\end{equation}
The primaries for a given model are labeled by a pair of integers $(r,s)$ where:
\begin{equation}
h_{r,s} = \frac{(mr-m's)^{2} - (m-m')^{2}}{4mm'}
\end{equation}
$$ 1\le r \le m'-1 \,,\quad 1\le s\le m-1 \,,\quad (r,s)\sim(m'-r,m-s)$$
Due to the above identifications we have a total of $(m-1)(m'-1)/2$ number of independent primaries. Hence this is the dimension of the SL(2,$\,\mathbb{Z}$) representation which will act on the characters $\chi_{r,s}$.

The $T,S$ matrices in this representation are given as:
\begin{equation}
\label{486}
\begin{split}
T_{rs,vw} &= \delta_{r,v}\delta_{s,w} \exp(2\pi i \left(h_{r,s} - \frac{c}{24} \right)) \\[2mm]
S_{rs,vw} &= \sqrt{\frac{8}{mm'}}(-1)^{1+sv+rw}\sin(\frac{\pi rvm}{m'})\sin(\frac{\pi swm'}{m})
\end{split}
\end{equation}

\section{Sage calculations for \sutk}

\label{sagecalc}

\begin{longtable}{|>{\centering}p{2.5cm}|>{\centering}p{5cm}|>{\centering}p{6.5cm}|}
\hline

 \hline
 \textbf{Level} $(k, c, N_{0})$ & \textbf{Seed primary label} $\lambda$  & \textbf{Poincar\'{e} sum}   \tabularnewline
 \hline
 \hline

$(1, 1, 4)$ & $1,2$  & $24\, Z_{A}$  \tabularnewline
\hline

$(2, \frac{3}{2}, 16)$ & $1,2,3$  & $32\, Z_{A}$ \tabularnewline
\hline

$(3, \frac{9}{5}, 20)$ & $1,2,3,4$  & $36\, Z_{A}$ \tabularnewline
\hline

$(4, 2, 20)$ & odd $,\ne 3$   & $8\, (4Z_{A} + Z_{D})$ \tabularnewline
		   & even  &  $32\, (2Z_{A} - Z_{D})$ \tabularnewline
		   & $3$  &  $48\, Z_{D}$ \tabularnewline
\hline

$(5, \frac{15}{7}, 28)$ & $1,\dots,6$ & $24\, Z_{A}$ \tabularnewline
\hline

$(6, \frac{9}{4}, 32)$ & $1,3,4,5,7$ & $32\, (Z_{A} + Z_{D})$ \tabularnewline
				    & $2,6$ & $16\, (7Z_{A} - 5Z_{D})$ \tabularnewline
\hline

$(7, \frac{7}{3}, 36)$ & $1,\dots,8$  & $54\, Z_{A}$ \tabularnewline
\hline


$(8, \frac{12}{5}, 40)$ & odd $,\ne 5$ & $24\, (2 Z_{A} + Z_{D})$ \tabularnewline
   				     & even  & $48\, (2 Z_{A} - Z_{D})$ \tabularnewline
   				     & $5$  & $96\, Z_{D}$ \tabularnewline
\hline

$(9, \frac{27}{11}, 44)$ & $1,\dots,10$  & $54\, Z_{A}$ \tabularnewline
\hline

$(10, \frac{5}{2}, 48)$ & $1,5,7,11$  & $32(Z_{A} + Z_{D} + Z_{E}^{(10)})$ \tabularnewline
					 & $2,10$  & $32(4Z_{A} - 2Z_{D} - Z_{E}^{(10)})$ \tabularnewline
					 & $3,6,9$  & $64(Z_{A} + Z_{D} - Z_{E}^{(10)})$ \tabularnewline
					 & $4,8$  & $32(3Z_{A} - 3Z_{D} + 2Z_{E}^{(10)})$ \tabularnewline
\hline


$(11, \frac{33}{13}, 52)$ & $1,\dots,12$   & $84\, Z_{A}$ \tabularnewline
\hline

$(12, \frac{18}{7}, 56)$ & odd $,\ne 7$   & $8\, (8Z_{A} + 5Z_{D})$ \tabularnewline
					  & even & $64\, (2Z_{A} - Z_{D})$ \tabularnewline
					  & $7$  & $144\, Z_{D}$ \tabularnewline
\hline

$(13, \frac{13}{5}, 60)$ & $1,2,4,7,8,11,13,14$   & $8\, (6Z_{A} + Z_{new}^{(13)})$ \tabularnewline
					  & $3,6,9,12$  & $72\, (2Z_{A} - Z_{new}^{(13)})$ \tabularnewline
					  & $5,10$  & $96\, Z_{new}^{(13)}$ \tabularnewline
\hline

$(14, \frac{21}{8}, 64)$ & $1,3,5,7,8,9,11,13,15$   & $64\, (Z_{A} + Z_{D})$ \tabularnewline
					  & $2,4,6,10,12,14$  & $32\, (5Z_{A} - 3Z_{D})$ \tabularnewline
\hline

$(15, \frac{45}{17}, 68)$ & $1,\dots,16$   & $108\, Z_{A}$ \tabularnewline

\hline
$(16, \frac{8}{3}, 72)$ & $1,5,7,11,13,17$   & $72\, (Z_{A} + Z_{E})$ \tabularnewline
					 & $2,4,6,8,10,12,14,16$  & $72\, (2Z_{A} - Z_{D})$ \tabularnewline
					 & $3,15$  & $24\, (3Z_{A} + 7Z_{D} - 7Z_{E})$ \tabularnewline
					 & $9$  & $48\, (5Z_{D} - 2Z_{E})$ \tabularnewline
\hline

$(17, \frac{51}{19}, 76)$ & $1,\dots,18$   & $120\, Z_{A}$ \tabularnewline

\hline

$(18, \frac{27}{10}, 80)$ & $1,3,7,9,11,13,17,19$   & $32\, (2Z_{A} + 2Z_{D} + Z_{new}^{(18)} )$ \tabularnewline
					   & $2,6,14,18$  & $16\, (13Z_{A} - 5Z_{D} - 4Z_{new}^{(18)} )$ \tabularnewline
					   & $4,8,12,16$  & $48\, (3Z_{A} - 3Z_{D} + 2Z_{new}^{(18)} )$ \tabularnewline
					   & $5,10,15$  & $128\, (Z_{A} + Z_{D} - Z_{new}^{(18)} )$ \tabularnewline
\hline

$(19, \frac{19}{7}, 84)$ & $\ne 0 \mbox{ mod } 3,7$   & $24\, (4Z_{A} +  Z_{new}^{(19)} )$ \tabularnewline
					   & $=0 \mbox{ mod } 3$  & $96\, (2Z_{A} - Z_{new}^{(19)} )$ \tabularnewline
					   & $=0 \mbox{ mod } 7$  & $144\, Z_{new}^{(19)} $ \tabularnewline
\hline


$(20, \frac{30}{11}, 88)$ & odd $,\ne 11$   & $24\, (4Z_{A} + 3Z_{D})$ \tabularnewline
					   & even  & $96\, (2Z_{A} - Z_{D})$ \tabularnewline
					   & $11$  & $240\, Z_{D}$ \tabularnewline
\hline


$(21, \frac{63}{23}, 92)$ & $1,\dots,22$   & $144\, Z_{A}$ \tabularnewline
\hline


$(22, \frac{11}{4}, 96)$ & $1,5,7,11,13,17,19,23$   & $32\, ( 2Z_{A} + 2Z_{D} + Z_{new1}^{(22)} )$ \tabularnewline
					  & $2,10,14,22$  & $16\, ( 10Z_{A} - 6Z_{D} + Z_{new1}^{(22)} + Z_{new2}^{(22)} )$ \tabularnewline
					  & $3,9,12,15,21$  & $64\, ( 2Z_{A} + 2Z_{D} - Z_{new1}^{(22)}  )$ \tabularnewline
					  & $4,20$  & $32\, ( 10Z_{A} - 6Z_{D} - Z_{new1}^{(22)} - 2Z_{new2}^{(2)} )$ \tabularnewline
					  & $6,18$  & $32\, ( 2Z_{A} + 2Z_{D} - Z_{new1}^{(22)} + 3Z_{new2}^{(2)} )$ \tabularnewline
					  & $8,16$  & $64\, ( 4Z_{A} - 4Z_{D} + Z_{new1}^{(22)} - Z_{new2}^{(2)} )$ \tabularnewline
\hline


$(23, \frac{69}{25}, 100)$ & $1,\dots,24$   & $150\, Z_{A}$ \tabularnewline
\hline


$(24, \frac{36	}{13}, 104)$ & odd $,\ne 13$   & $8\, (14Z_{A} + 11Z_{D})$ \tabularnewline
						& even  & $112\, (2Z_{A} - Z_{D})$ \tabularnewline
						& $13$  & $288\, Z_{D}$ \tabularnewline
\hline


$(25, \frac{25}{9}, 108)$ & $\ne 0 \mbox{ mod } 3$   & $54\, (3Z_{A} - Z_{new}^{(25)})$ \tabularnewline
					   & $=0 \mbox{ mod } 3$  & $36\, (3Z_{A} + 5Z_{new}^{(25)})$ \tabularnewline
\hline

$(26, \frac{39}{14}, 112)$ & $\ne 0 \mbox{ mod } 2,7 $   & $32\, (3Z_{A} + 3Z_{D} + Z_{new}^{(26)})$ \tabularnewline
					    & $=2 \mbox{ mod } 4 ,\ne 14 $  & $96\, (3Z_{A} - Z_{D} - Z_{new}^{(26)})$ \tabularnewline
					    & $=0 \mbox{ mod } 4  $  & $64\, (3Z_{A} - 3Z_{D} + 2Z_{new}^{(26)})$ \tabularnewline
					    & $=0 \mbox{ mod } 7  $  & $192\, (Z_{A} + Z_{D} - Z_{new}^{(26)})$ \tabularnewline
\hline


$(27, \frac{81}{29}, 116)$ & $1,\dots,28$   & $180\, Z_{A}$ \tabularnewline
\hline


$(28, \frac{14}{5}, 120)$ & $1,7,11,13,17,19,23,29$   & $8\,( 12Z_{A} + 8Z_{D} + 11Z_{E} + Z_{new}^{(28)})$ \tabularnewline
					   & $2,4,8,14,16,22,26,28$  & $16\,( 12Z_{A} - 4Z_{D} - Z_{E} + Z_{new}^{(28)})$ \tabularnewline
					   & $3,9,21,27$  & $48\,( 4Z_{A} - Z_{E} - Z_{new}^{(28)})$ \tabularnewline
					   & $5,25$  & $32\,( 10Z_{D} - 5Z_{E} + 2Z_{new}^{(28)})$ \tabularnewline
					   & $6,12,18,24$  & $96\,( 4Z_{A} - 4Z_{D} + Z_{E} - Z_{new}^{(28)})$ \tabularnewline
					   & $10,20$  & $128\,( 2Z_{D} - Z_{E} + Z_{new}^{(28)})$ \tabularnewline
					   & $15$  & $192\,( 2Z_{D} - Z_{E})$ \tabularnewline
\hline


$(29, \frac{87}{31}, 124)$ & $1,\dots,30$   & $192\, Z_{A}$ \tabularnewline
\hline


$(30, \frac{45}{16}, 128)$ & odd, $16$   & $128\, (Z_{A} + Z_{D})$ \tabularnewline
						& even, $\ne 4,12,16,20,28$  & $64\, (5Z_{A} - 3Z_{D} - 2Z_{new}^{(30)})$ \tabularnewline
						& $4,12,20,28$  & $64\, (3Z_{A} - Z_{D} + 5Z_{new}^{(30)})$ \tabularnewline
\hline

$(31, \frac{31}{11}, 132)$ & $\ne 0 \mbox{ mod } 3,11$   & $48\, (3Z_{A} + Z_{new}^{(31)})$ \tabularnewline
						& $=0 \mbox{ mod } 3$  & $144\, (2Z_{A} - Z_{new}^{(31)})$ \tabularnewline
						& $=0 \mbox{ mod } 11$  & $240\, Z_{new}^{(31)}$ \tabularnewline

\hline

$(32, \frac{48}{17}, 136)$ & odd $,\ne 17$  & $24\, (6Z_{A} + 5Z_{D}) $ \tabularnewline
						& even  & $144\, (2Z_{A} - Z_{D}) $ \tabularnewline
						& $17$  & $384\, Z_{D} $ \tabularnewline
\hline


\hline
$(33, \frac{99}{35}, 140)$ & $\ne 0 \mbox{ mod } 5,7$   & $12\, (16Z_{A} + Z_{new}^{(33)}) $ \tabularnewline
						& $=0 \mbox{ mod } 5$  & $192\, (2Z_{A} - Z_{new}^{(33)}) $ \tabularnewline
						& $=0 \mbox{ mod } 7$  & $216\, Z_{new}^{(33)} $ \tabularnewline
\hline

$(34, \frac{17}{6}, 144)$ & odd, $\ne 3,9,15,21,27,33$   & $48\, (3Z_{A} + 3Z_{D} -2Z_{new1}^{(34)})$ \tabularnewline
					   & $2,10,14,22,26,34$  & $24\, (9Z_{A} - 9Z_{D} - 2Z_{new1}^{(34)} + 2Z_{new2}^{(34)})$ \tabularnewline
					   & $3,15,21,33$  & $16\, (9Z_{A} + 9Z_{D} + 12Z_{new1}^{(34)} - 2Z_{new2}^{(34)})$ \tabularnewline
					   & $4,8,12,16,20,24,28,32$  & $24\, (15Z_{A} - 3Z_{D} - 2Z_{new1}^{(34)} - 2Z_{new2}^{(34)})$ \tabularnewline
					   & $6,30$  & $8\, (27Z_{A} - 27Z_{D} + 30Z_{new1}^{(34)} + 2Z_{new2}^{(34)})$ \tabularnewline
					   & $9,18,27$  & $64\, (3Z_{new1}^{(34)} + Z_{new2}^{(34)})$ \tabularnewline
\hline

$(35, \frac{105}{37}, 148)$ & $1,\dots,36$   & $228\, Z_{A}$ \tabularnewline
\hline


$(36, \frac{54}{19}, 152)$ & odd $,\ne 19$   & $ 8\, (20Z_{A} + 17Z_{D})$ \tabularnewline
						& even  & $ 160\, (2Z_{A} - Z_{D})$ \tabularnewline
						& $19$  & $ 432\, Z_{D}$ \tabularnewline
\hline


$(37, \frac{37}{13}, 156)$ & $\ne0 \mbox{ mod } 3,13$   & $12\, (14Z_{A} + 5Z_{new}^{(37)})$ \tabularnewline
						& $=0 \mbox{ mod } 3$  & $168\, (2Z_{A} - Z_{new}^{(37)})$ \tabularnewline
						& $=0 \mbox{ mod } 13$  & $288\, Z_{new}^{(37)}$ \tabularnewline
\hline


$(38, \frac{57}{20}, 160)$ & odd, $\ne 5,15,25,35$ & $32\, ( 4Z_{A} + 4Z_{D} + Z_{new1}^{(38)})$ \tabularnewline
						& $2,6,14,18,22,26,34,38$  & $16\, (16Z_{A} - 8Z_{D} + Z_{new1}^{(38)} + 6Z_{new2}^{(38)} )$ \tabularnewline
						& $4,12,28,36$  & $64\, (8Z_{A} - 4Z_{D} - Z_{new1}^{(38)} - 3Z_{new2}^{(38)} )$ \tabularnewline
						& $5,15,20,25,35$  & $128\, (2Z_{A} + 2Z_{D} - Z_{new1}^{(38)} )$ \tabularnewline
						& $8,16,24,32$  & $96\, (4Z_{A} - 4Z_{D} + Z_{new1}^{(38)} - 2Z_{new2}^{(38)} )$ \tabularnewline
						& $10,30$  & $64\, (2Z_{A} + 2Z_{D} - Z_{new1}^{(38)} + 6Z_{new2}^{(38)} )$ \tabularnewline
\hline

$(39, \frac{117}{41}, 164)$ & $1,\dots,40$   & $252\, Z_{A}$ \tabularnewline
\hline


$(40, \frac{20}{7}, 168)$ & odd, $\ne 3 \mbox{ mod }6 , \ne 7,35$    & $8\, ( 16Z_{A} + 40Z_{D} - 15Z_{new1}^{(40)} + 2Z_{new2}^{(40)} )$ \tabularnewline
					   & even, $\ne 0 \mbox{ mod } 6 ,\ne 14,28$   & $32\, ( 8Z_{A} - 4Z_{D} + Z_{new2}^{(40)})$ \tabularnewline
					   & $3,9,15,27,33,39$  & $16\, ( 16Z_{A} - 8Z_{D} + 9Z_{new1}^{(40)} - 4Z_{new2}^{(40)})$ \tabularnewline
					   & $6,12,18,24,30,36$  & $128\, (4Z_{A} - 2Z_{D} - Z_{new2}^{(40)} )$ \tabularnewline
					   & $7,35$  & $48\, ( 3Z_{new1}^{(40)} + 2Z_{new2}^{(40)} )$ \tabularnewline					   
					   & $14,28$  & $192\, Z_{new2}^{(40)}$ \tabularnewline
					   & $21$  & $288\, Z_{new1}^{(40)}$ \tabularnewline
\hline


$(41, \frac{123}{43}, 172)$ & $1,\dots,42$   & $264\, Z_{A}$ \tabularnewline
\hline

\caption{\poin sums for \sutk}
\label{SU2fulltable}
\end{longtable}

In the above table, terms labelled $Z_{new}$ are unphysical modular invariants. We list a few of them below. Notice that for $k=22$ there are two unphysical invariants. This number grows for generic $k$. 
\be
\begin{split}
Z^{(13)}_{new} &= \abs{\chi_{1} - \chi_{11}}^{2} + \abs{\chi_{2} + \chi_{8}}^{2} + \abs{\chi_{4} - \chi_{14}}^{2} + \abs{\chi_{7} + \chi_{13}}^{2} + 2\abs{\chi_{5}}^{2} + 2\abs{\chi_{10}}^{2} \\
\\
Z^{(19)}_{new} &= \abs{\chi_{1} + \chi_{13}}^{2} + \abs{\chi_{2} - \chi_{16}}^{2} + \abs{\chi_{4} + \chi_{10}}^{2} + \abs{\chi_{5} - \chi_{19}}^{2} + \abs{\chi_{8} + \chi_{20}}^{2} + \abs{\chi_{11} + \chi_{17}}^{2} \\
&\quad + 2\abs{\chi_{7}}^{2} + 2\abs{\chi_{14}}^{2} \\
\\
Z_{new}^{(18)} &= \abs{\chi_{1} - \chi_{9}}^{2} + \abs{\chi_{3} + \chi_{13}}^{2} + \abs{\chi_{4} + \chi_{16}}^{2} + \abs{\chi_{7} + \chi_{17}}^{2} + \abs{\chi_{8} + \chi_{12}}^{2} + \abs{\chi_{11} - \chi_{19}}^{2} \\
\\
Z_{new1}^{(22)} &= 2\left( \abs{\chi_{1} - \chi_{17}}^{2} + \abs{\chi_{5} + \chi_{11}}^{2} + \abs{\chi_{7} - \chi_{23}}^{2} + \abs{\chi_{13} + \chi_{19}}^{2} + \abs{\chi_{8} + \chi_{16}}^{2} \right) \\
&\quad + \abs{\chi_{2} + \chi_{10} + \chi_{14} + \chi_{22}}^{2}\\
Z_{new2}^{(22)} &= 4 \abs{\chi_{6} - \chi_{18}}^{2} + 2\abs{\chi_{2} + \chi_{10} - \chi_{14} - \chi_{22}}^{2} \\
\end{split}
\ee

\section{Sage calculations for unitary Virasoro minimal models}

\begin{longtable}{|>{\centering}p{3cm}|>{\centering}p{3.5cm}|>{\centering}p{3cm}|>{\centering}p{5.5cm}| }
 \hline
 \textbf{Minimal Model} $(m,m',c,N_{0})$ & \textbf{Seed Primary} $(\chi_{r,s})$ & \textbf{Dimension} $(h_{r,s})$ & \textbf{Partition function}  \tabularnewline
 \hline
 \hline
	Ising               & $\chi_{1,1} , \, \chi_{1,2}$ & $0, \frac{1}{2}$  & $32\, Z_{AA}$ \tabularnewline
	$(3,4, \frac{1}{2},16)$  &  $\chi_{2,2}$ & $\frac{1}{16}$  & $32\, Z_{AA}$  \tabularnewline
 \hline	            
	Tricritical Ising      & $\chi_{1,1} , \, \chi_{1,3} , \, \chi_{2,3} , \, \chi_{3,3}$ & $0, \frac{3}{2}, \frac{3}{5}, \frac{1}{10}$  & $384\, Z_{AA}$  \tabularnewline       
	$(4,5,\frac{7}{10},80)$  &  $\chi_{1,2} , \, \chi_{2,2}$ & $\frac{7}{16}, \frac{3}{80}$& $384\, Z_{AA}$  \tabularnewline 
 \hline
    3-State Potts    & $\chi_{1,1} , \, \chi_{1,2} , \, \chi_{1,3} , \, \chi_{1,4}$ & $0, \frac{2}{5}, \frac{7}{5}, 3$ & $96( 4Z_{AA} + Z_{AD} )$  \tabularnewline
	$(5,6,\frac{4}{5},120)$	  & $\chi_{3,3} , \, \chi_{3,4}$ & $\frac{1}{15}, \frac{2}{3}$  & $576 \, Z_{AD} $  \tabularnewline
    				  &  $\chi_{2,2} , \, \chi_{2,3} , \, \chi_{2,4} , \, \chi_{4,4} $ & $\frac{1}{40}, \frac{21}{40}, \frac{13}{8}, \frac{1}{8}$  & $384( 2Z_{AA} - Z_{AD} )$  \tabularnewline
 \hline
 Tricritical Potts            & $\chi_{1,1}$  & $0$  & $128(4Z_{AA} + Z_{AD})$ \tabularnewline
 $(6,7,\frac{6}{7},168)$	  & $\chi_{1,3}$ & $\frac{4}{3}$ & $768 \, Z_{AD}$ \tabularnewline
 					      & $\chi_{1,2}$ & $\frac{3}{8}$ & $512(2Z_{AA} - Z_{AD})$ \tabularnewline
\hline

 Name				      & $\chi_{1,1}$ & $0$ &  $512(Z_{AA} + Z_{AD})$ \tabularnewline
 $(7,8,\frac{25}{28},224)$	  & $\chi_{4,4}$ & $\frac{15}{224}$ &  $512(Z_{AA} + Z_{AD})$ \tabularnewline
 					      & $\chi_{2,2}$ & $\frac{3}{224}$ &  $256(7\,Z_{AA} - 5\,Z_{AD})$ \tabularnewline
 \hline
 Name				      & $\chi_{1,1}$ & $0$ &  $572(Z_{AA} + Z_{AD})$ \tabularnewline
 $(8,9,\frac{11}{12},288)$	  & $\chi_{1,4}$ & $\frac{87}{32}$ &  $572(Z_{AA} + Z_{AD})$ \tabularnewline
 					  	  & $\chi_{1,2}$ & $\frac{11}{32}$ &  $288(7\,Z_{AA} - 5\,Z_{AD})$ \tabularnewline
 					  	  & $\chi_{3,3}$ & $\frac{1}{36}$ &  $572(Z_{AA} + Z_{AD})$ \tabularnewline
 					   	  & $\chi_{3,4}$ & $\frac{143}{288}$ &  $572(Z_{AA} + Z_{AD})$ \tabularnewline
 					  	  & $\chi_{3,6}$ & $\frac{899}{288}$ &  $288(7\,Z_{AA} - 5\,Z_{AD})$ \tabularnewline

\hline

Name				  	  & $\chi_{1,1}$ & $0$  & $432(2Z_{AA} + Z_{AD})$ \tabularnewline
$(9,10,\frac{11}{12},360)$	  & $\chi_{2,2}$ & $\frac{1}{120}$  & $864(2Z_{AA} - Z_{AD})$ \tabularnewline
 						  & $\chi_{5,6}$ & $\frac{28}{45}$ &  $1728Z_{AD}$ \tabularnewline 						  
 						  & $\chi_{1,3}$ & $\frac{11}{9}$ & $432(2Z_{AA} + Z_{AD})$ \tabularnewline
 					  	  & $\chi_{2,3}$ & $\frac{143}{360}$ &  $864(2Z_{AA} - Z_{AD})$ \tabularnewline
 					  	  & $\chi_{5,5}$ & $\frac{1}{15}$ &  $1728\,Z_{AD}$ \tabularnewline

\hline

Name		              & $\chi_{1,1}$  & $0$  & $576(2Z_{AA} + Z_{AD})$  \tabularnewline
$(10,11,\frac{52}{55},440)$	  & $\chi_{1,2}$ & $\frac{13}{40}$  & $1152(2Z_{AA} - Z_{AD})$  \tabularnewline
 					      & $\chi_{1,5}$ & $\frac{23}{5}$  & $2304\, Z_{AD}$  \tabularnewline

\hline

Name		              & $\chi_{1,1}$  & $0$ &  $768(Z_{AA} + Z_{AD} + Z^{(12)})$  \tabularnewline
$(11,12,\frac{21}{22},528)$	  & $\chi_{2,2}$ & $\frac{1}{176}$ &  $768(4 Z_{AA} - 2 Z_{AD} - Z^{(12)})$  \tabularnewline
 					      & $\chi_{3,3}$ & $\frac{1}{66}$ &  $1536(Z_{AA} + Z_{AD} - Z^{(12)})$  \tabularnewline
						  & $\chi_{4,4}$  & $\frac{5}{176}$ &  $768(3Z_{AA} - 3Z_{AD} + 2Z^{(12)})$  \tabularnewline
						  & $\chi_{6,6}$  & $\frac{35}{528}$ &  $1536(Z_{AA} + Z_{AD} - Z^{(12)})$ \tabularnewline
\hline

\caption{Poincar\'{e} Sums for unitary Virasoro minimal models}
\label{virmin}
\end{longtable}

\bibliographystyle{JHEP}
\bibliography{Poincare}

\providecommand{\href}[2]{#2}\begingroup\raggedright\begin{thebibliography}{10}

\bibitem{Maloney:2007ud}
A.~Maloney and E.~Witten, \emph{{Quantum Gravity Partition Functions in Three
  Dimensions}}, \href{https://doi.org/10.1007/JHEP02(2010)029}{\emph{JHEP}
  {\bfseries 02} (2010) 029} [\href{https://arxiv.org/abs/0712.0155}{{\ttfamily
  0712.0155}}].

\bibitem{Dijkgraaf:2000fq}
R.~Dijkgraaf, J.~M. Maldacena, G.~W. Moore and E.~P. Verlinde, \emph{{A Black
  Hole Farey Tail}},  \href{https://arxiv.org/abs/hep-th/0005003}{{\ttfamily
  hep-th/0005003}}.

\bibitem{Manschot:2007zb}
J.~Manschot, \emph{{AdS(3) Partition Functions Reconstructed}},
  \href{https://doi.org/10.1088/1126-6708/2007/10/103}{\emph{JHEP} {\bfseries
  10} (2007) 103} [\href{https://arxiv.org/abs/0707.1159}{{\ttfamily
  0707.1159}}].

\bibitem{Manschot:2010mod}
J.~Manschot and G.~W. Moore, \emph{{A Modern Fareytail}}, {\emph{Communications
  in Number Theory and Physics} {\bfseries 4} (2010) 103}.

\bibitem{Giombi:2008vd}
S.~Giombi, A.~Maloney and X.~Yin, \emph{{One-loop Partition Functions of 3D
  Gravity}}, \href{https://doi.org/10.1088/1126-6708/2008/08/007}{\emph{JHEP}
  {\bfseries 08} (2008) 007} [\href{https://arxiv.org/abs/0804.1773}{{\ttfamily
  0804.1773}}].

\bibitem{Keller:2014xba}
C.~A. Keller and A.~Maloney, \emph{{Poincare Series, 3D Gravity and CFT
  Spectroscopy}}, \href{https://doi.org/10.1007/JHEP02(2015)080}{\emph{JHEP}
  {\bfseries 02} (2015) 080} [\href{https://arxiv.org/abs/1407.6008}{{\ttfamily
  1407.6008}}].

\bibitem{Castro:2011zq}
A.~Castro, M.~R. Gaberdiel, T.~Hartman, A.~Maloney and R.~Volpato, \emph{{The
  Gravity Dual of the Ising Model}},
  \href{https://doi.org/10.1103/PhysRevD.85.024032}{\emph{Phys. Rev.}
  {\bfseries D85} (2012) 024032}
  [\href{https://arxiv.org/abs/1111.1987}{{\ttfamily 1111.1987}}].

\bibitem{Jian:2019ubz}
C.-M. Jian, A.~W.~W. Ludwig, Z.-X. Luo, H.-Y. Sun and Z.~Wang,
  \emph{{Establishing strongly-coupled 3D AdS quantum gravity with Ising dual
  using all-genus partition functions}},
  \href{https://doi.org/10.1007/JHEP10(2020)129}{\emph{JHEP} {\bfseries 10}
  (2020) 129} [\href{https://arxiv.org/abs/1907.06656}{{\ttfamily
  1907.06656}}].

\bibitem{Saad:2019lba}
P.~Saad, S.~H. Shenker and D.~Stanford, \emph{{JT gravity as a matrix
  integral}},  \href{https://arxiv.org/abs/1903.11115}{{\ttfamily 1903.11115}}.

\bibitem{Witten:2020wvy}
E.~Witten, \emph{{Matrix Models and Deformations of JT Gravity}},
  \href{https://doi.org/10.1098/rspa.2020.0582}{\emph{Proc. Roy. Soc. Lond. A}
  {\bfseries 476} (2020) 20200582}
  [\href{https://arxiv.org/abs/2006.13414}{{\ttfamily 2006.13414}}].

\bibitem{Maloney:2020nni}
A.~Maloney and E.~Witten, \emph{{Averaging over Narain moduli space}},
  \href{https://doi.org/10.1007/JHEP10(2020)187}{\emph{JHEP} {\bfseries 10}
  (2020) 187} [\href{https://arxiv.org/abs/2006.04855}{{\ttfamily
  2006.04855}}].

\bibitem{Afkhami-Jeddi:2020ezh}
N.~Afkhami-Jeddi, H.~Cohn, T.~Hartman and A.~Tajdini, \emph{{Free partition
  functions and an averaged holographic duality}},
  \href{https://arxiv.org/abs/2006.04839}{{\ttfamily 2006.04839}}.

\bibitem{Bantay:2001ni}
P.~Bantay, \emph{{The kernel of the modular representation and the Galois
  action in RCFT}},
  \href{https://doi.org/10.1007/s00220-002-0760-x}{\emph{Commun. Math. Phys.}
  {\bfseries 233} (2003) 423}
  [\href{https://arxiv.org/abs/math/0102149}{{\ttfamily math/0102149}}].

\bibitem{Saad:2018bqo}
P.~Saad, S.~H. Shenker and D.~Stanford, \emph{{A semiclassical ramp in SYK and
  in gravity}},  \href{https://arxiv.org/abs/1806.06840}{{\ttfamily
  1806.06840}}.

\bibitem{Cotler:2020ugk}
J.~Cotler and K.~Jensen, \emph{{AdS$_3$ gravity and random CFT}},
  \href{https://arxiv.org/abs/2006.08648}{{\ttfamily 2006.08648}}.

\bibitem{Cotler:2020hgz}
J.~Cotler and K.~Jensen, \emph{{AdS$_3$ wormholes from a modular bootstrap}},
  \href{https://doi.org/10.1007/JHEP11(2020)058}{\emph{JHEP} {\bfseries 11}
  (2020) 058} [\href{https://arxiv.org/abs/2007.15653}{{\ttfamily
  2007.15653}}].

\bibitem{Benjamin:2019stq}
N.~Benjamin, H.~Ooguri, S.-H. Shao and Y.~Wang, \emph{{Light-cone modular
  bootstrap and pure gravity}},
  \href{https://doi.org/10.1103/PhysRevD.100.066029}{\emph{Phys. Rev. D}
  {\bfseries 100} (2019) 066029}
  [\href{https://arxiv.org/abs/1906.04184}{{\ttfamily 1906.04184}}].

\bibitem{Benjamin:2020mfz}
N.~Benjamin, S.~Collier and A.~Maloney, \emph{{Pure Gravity and Conical
  Defects}}, \href{https://doi.org/10.1007/JHEP09(2020)034}{\emph{JHEP}
  {\bfseries 09} (2020) 034}
  [\href{https://arxiv.org/abs/2004.14428}{{\ttfamily 2004.14428}}].

\bibitem{Alday:2019vdr}
L.~F. Alday and J.-B. Bae, \emph{{Rademacher Expansions and the Spectrum of 2d
  CFT}}, \href{https://doi.org/10.1007/JHEP11(2020)134}{\emph{JHEP} {\bfseries
  11} (2020) 134} [\href{https://arxiv.org/abs/2001.00022}{{\ttfamily
  2001.00022}}].

\bibitem{Maxfield:2020ale}
H.~Maxfield and G.~J. Turiaci, \emph{{The path integral of 3D gravity near
  extremality; or, JT gravity with defects as a matrix integral}},
  \href{https://arxiv.org/abs/2006.11317}{{\ttfamily 2006.11317}}.

\bibitem{Cappelli:1986hf}
A.~Cappelli, C.~Itzykson and J.~B. Zuber, \emph{{Modular Invariant Partition
  Functions in Two-Dimensions}},
  \href{https://doi.org/10.1016/0550-3213(87)90155-6}{\emph{Nucl. Phys.}
  {\bfseries B280} (1987) 445}.

\bibitem{Cappelli:1987xt}
A.~Cappelli, C.~Itzykson and J.~B. Zuber, \emph{{The ADE Classification of
  Minimal and A1(1) Conformal Invariant Theories}},
  \href{https://doi.org/10.1007/BF01221394}{\emph{Commun. Math. Phys.}
  {\bfseries 113} (1987) 1}.

\bibitem{Chandra:2018pjq}
A.~R. Chandra and S.~Mukhi, \emph{{Towards a Classification of Two-Character
  Rational Conformal Field Theories}},
  \href{https://doi.org/10.1007/JHEP04(2019)153}{\emph{JHEP} {\bfseries 04}
  (2019) 153} [\href{https://arxiv.org/abs/1810.09472}{{\ttfamily
  1810.09472}}].

\bibitem{Chandra:2018ezv}
A.~R. Chandra and S.~Mukhi, \emph{{Curiosities above c = 24}},
  \href{https://doi.org/10.21468/SciPostPhys.6.5.053}{\emph{SciPost Phys.}
  {\bfseries 6} (2019) 053} [\href{https://arxiv.org/abs/1812.05109}{{\ttfamily
  1812.05109}}].

\bibitem{Mukhi:2020gnj}
S.~Mukhi, R.~Poddar and P.~Singh, \emph{{Rational CFT with three characters:
  the quasi-character approach}},
  \href{https://doi.org/10.1007/JHEP05(2020)003}{\emph{JHEP} {\bfseries 05}
  (2020) 003} [\href{https://arxiv.org/abs/2002.01949}{{\ttfamily
  2002.01949}}].

\bibitem{Bagger:2012jb}
J.~Bagger, N.~Lambert, S.~Mukhi and C.~Papageorgakis, \emph{{Multiple Membranes
  in M-theory}},
  \href{https://doi.org/10.1016/j.physrep.2013.01.006}{\emph{Phys. Rept.}
  {\bfseries 527} (2013) 1} [\href{https://arxiv.org/abs/1203.3546}{{\ttfamily
  1203.3546}}].

\bibitem{Sage}
W.~Stein et~al., \emph{{S}age {M}athematics {S}oftware ({V}ersion 2020.2)}.
\newblock The Sage Development Team, 2020.

\bibitem{Gannon:1992ty}
T.~Gannon, \emph{{The Classification of affine SU(3) modular invariant
  partition functions}},
  \href{https://doi.org/10.1007/BF02099776}{\emph{Commun. Math. Phys.}
  {\bfseries 161} (1994) 233}
  [\href{https://arxiv.org/abs/hep-th/9212060}{{\ttfamily hep-th/9212060}}].

\bibitem{Gannon:1994cf}
T.~Gannon, \emph{{The Classification of SU(3) modular invariants revisited}},
  {\emph{Ann. Inst. H. Poincare Phys. Theor.} {\bfseries 65} (1996) 15}
  [\href{https://arxiv.org/abs/hep-th/9404185}{{\ttfamily hep-th/9404185}}].

\bibitem{Itzykson:1988hk}
C.~Itzykson, \emph{{Level One Kac-Moody Characters And Modular Invariance}},
  in \emph{Annecy 1988, Proceedings, Conformal Field Theories And Related
  Topics}, 1988.

\bibitem{Degiovanni:1989ne}
P.~Degiovanni, \emph{{Z/NZ Conformal Field Theories}},
  \href{https://doi.org/10.1007/BF02096494}{\emph{Commun. Math. Phys.}
  {\bfseries 127} (1990) 71}.

\bibitem{Bauer:1990xx}
M.~{Bauer} and C.~{Itzykson}, \emph{{Modular transformations of SU(N) affine
  characters and their commutant}},
  \href{https://doi.org/10.1007/BF02104506}{\emph{Commun. Math. Phys.}
  {\bfseries 127} (1990) 617}.

\bibitem{Schellekens:1992db}
A.~N. Schellekens, \emph{{Meromorphic c = 24 conformal field theories}},
  \href{https://doi.org/10.1007/BF02099044}{\emph{Commun. Math. Phys.}
  {\bfseries 153} (1993) 159}
  [\href{https://arxiv.org/abs/hep-th/9205072}{{\ttfamily hep-th/9205072}}].

\bibitem{Goddard:1983at}
P.~Goddard and D.~I. Olive, \emph{{Algebras, Lattices and Strings}},  in
  \emph{{Goddard, P. (Ed.), Olive, D. (Ed.): Ka-Moody And Virasoro Algebras*
  210-255}}, 1983.

\bibitem{Karch:2020flx}
A.~Karch, Z.-X. Luo and H.-Y. Sun, \emph{{Holographic duality for Ising CFT
  with boundary}},  \href{https://arxiv.org/abs/2012.02067}{{\ttfamily
  2012.02067}}.

\bibitem{Randall:1999vf}
L.~Randall and R.~Sundrum, \emph{{An Alternative to compactification}},
  \href{https://doi.org/10.1103/PhysRevLett.83.4690}{\emph{Phys. Rev. Lett.}
  {\bfseries 83} (1999) 4690}
  [\href{https://arxiv.org/abs/hep-th/9906064}{{\ttfamily hep-th/9906064}}].

\bibitem{Cardy:1986gw}
J.~L. Cardy, \emph{{Effect of Boundary Conditions on the Operator Content of
  Two-Dimensional Conformally Invariant Theories}},
  \href{https://doi.org/10.1016/0550-3213(86)90596-1}{\emph{Nucl. Phys. B}
  {\bfseries 275} (1986) 200}.

\bibitem{Mathur:1988na}
S.~D. Mathur, S.~Mukhi and A.~Sen, \emph{{On the Classification of Rational
  Conformal Field Theories}},
  \href{https://doi.org/10.1016/0370-2693(88)91765-0}{\emph{Phys. Lett.}
  {\bfseries B213} (1988) 303}.

\bibitem{Mukhi:2019xjy}
S.~Mukhi, \emph{{Classification of RCFT from Holomorphic Modular Bootstrap: A
  Status Report}},  in \emph{{Pollica Summer Workshop 2019: Mathematical and
  Geometric Tools for Conformal Field Theories (PSW2019) Pollica, Salerno,
  Italy, June 3-21, 2019}}, 2019,
  \href{https://arxiv.org/abs/1910.02973}{{\ttfamily 1910.02973}}.

\end{thebibliography}\endgroup

\end{document}